\documentclass[preprint,12pt,authoryear]{elsarticle}

    \usepackage{amssymb}
    \usepackage{amsmath}
    \usepackage{algorithm}
    \usepackage[noend]{algpseudocode}
    \usepackage{subcaption}
    \usepackage[english]{babel}
    \usepackage{xcolor}
    \usepackage{paralist}	
    \usepackage[lowtilde]{url}
    \usepackage{fixltx2e}
    \usepackage{listings}
    \usepackage{color}
    \usepackage[hidelinks]{hyperref}
    \usepackage{graphicx}
    \usepackage{geometry}
    \usepackage{mathabx}
    \usepackage{pst-all}
    \usepackage{pstricks-add}
    \usepackage{float}
    \usepackage{kantlipsum}
    \usepackage{dsfont}
    \usepackage{amssymb}
    \usepackage{multirow}
    \allowdisplaybreaks

\journal{Computational Statistics $\&$ Data Analysis}

\newcommand{\diag}{\mbox{{\bf diag}}}

\def\*#1{\boldsymbol{#1}}
\def\~#1{{\cal #1}}

\begin{document}

\begin{frontmatter}

\title{Model-based edge clustering for weighted networks with a noise component}

\author[inst1]{Haomin Li}

\affiliation[inst1]{organization={Department of Biostatistics},
            addressline={University of Iowa}, 
            city={Iowa City},
            postcode={52242}, 
            state={IA},
            country={USA}}

\author[inst1]{Daniel K. Sewell}
\ead{daniel-sewell@uiowa.edu}

\begin{abstract}
Clustering is a fundamental task in network analysis, essential for uncovering hidden structures within complex systems. Edge clustering, which focuses on relationships between nodes rather than the nodes themselves, has gained increased attention in recent years. However, existing edge clustering algorithms often overlook the significance of edge weights, which can represent the strength or capacity of connections, and fail to account for noisy edges—connections that obscure the true structure of the network. To address these challenges, the Weighted Edge Clustering Adjusting for Noise (WECAN) model is introduced. This novel algorithm integrates edge weights into the clustering process and includes a noise component that filters out spurious edges. WECAN offers a data-driven approach to distinguishing between meaningful and noisy edges, avoiding the arbitrary thresholding commonly used in network analysis. Its effectiveness is demonstrated through simulation studies and applications to real-world datasets, showing significant improvements over traditional clustering methods. Additionally, the R package "WECAN" \footnote{\url{https://github.com/HaominLi7/WECAN}} has been developed to facilitate its practical implementation.
\end{abstract}


\end{frontmatter}


\section{Introduction}
\label{sec:sample1}

Networks arise as dominant structures in various fields, including epidemiology, sociology, biology, neuroscience, and computer science. In network analysis, entities of interest are depicted as nodes, while the connections between them are typically denoted by edges. These edges can take the form of unordered pairs, indicating undirected relationships, or ordered pairs, signifying directed relationships \citep{butts2009revisiting}. One of the pivotal research areas within network analysis is clustering, of which community detection is a special case. Communities in a network are typically characterized by dense groups of nodes, exhibiting dense internal connections while maintaining sparser connections with nodes outside the group, although more broadly clusters of nodes tend to exhibit structural similarities. Network clustering serves as a cornerstone in unraveling the functionality of complex networks \citep{khan2017network}.

As one of the most broadly-studied topics in network science, various algorithms and statistical models have been developed for community detection. Among them, there are algorithmic methods such as the Kernighan-Lin algorithm for graph partitioning \citep{kernighan1970efficient}, the Girvan Newman (GN) algorithm \citep{newman2004finding}, the spectral clustering algorithm \citep{ng2001spectral}; in addition, many model-based methods exist such as the latent position cluster model \citep{handcock2007model}, and the stochastic block model \citep{nowicki2001estimation} and their many variants. 
These algorithms have been widely accepted and used in various fields. However, despite advantages in interpretation and intuition in many settings, all of these models focus on grouping nodes into clusters based on similarity, but ignore the potential that lay in edge clustering. 

The idea of edge-centric clustering is attractive through the following aspects. First, it is common for communities to be built based on the identities of the edges in realistic networks. For example, in a scientific collaboration network, nodes represent researchers and edges represent joint research efforts or co-authorship of papers. In such a network, communities emerge based on research topics, academic disciplines, or shared research interests. Second, membership in multiple organizations is extremely common. Through notions of mixed-membership, node-centric network clustering methods have been developed to address this; yet difficulties remain in interpreting the results, as a node can be divided into multiple proportions, each of which belongs to a distinct community \citep{amelio2014overlapping}. For instance, a person may be a member of multiple interest groups, yet it is counter-intuitive to say that $50\%$ of that person belongs to a dance club. Clustering on edges overcomes this issue, as edges represent relationships, and in most cases a relationship forms within a single context/cluster. 

Despite the advantages above, very little work has been devoted to the area of edge clustering when compared with the numerous node-centric clustering algorithms. A few algorithms tried to solve this question by performing standard community detection on the line graph \citep{evans2009line};  \citep{evans2010line}; \citep{yoshida2013weighted}; \citep{tian2023mixed}. In recent years, the latent space edge clustering (LSEC) model \citep{sewell2021model} was introduced as the first model-based edge community detection algorithm for directed networks. Building on this, the automated latent space edge clustering (aLSEC) model \citep{pham2024automated} was developed, using an overfitted mixture prior to automatically determine the number of clusters. This addresses the limitation of the LSEC model, which requires the number of clusters to be specified in advance. Even though the LSEC model and aLSEC model have been shown to be accurate and computationally efficient, there are two fundamental limitations, described below. 

The first limitation is that none of these edge clustering algorithms takes edge weights into consideration. Incorporating edge weights into the clustering model is crucial as they quantify the strength of relationships between connected nodes. Strongly weighted edges often signify more substantial interactions, indicative of specific communities or functional modules \citep{opsahl2009clustering}. While node-centric clustering algorithms, such as weighted stochastic blockmodels \citep{aicher2015learning}, have integrated edge weights to enhance accuracy, edge-centric clustering methods have largely overlooked this factor. In this study, we aim to fill this gap by developing the first model-based edge clustering algorithm tailored for weighted networks.

Second, while noise contamination has long been recognized as an important consideration in clustering of traditional data \citep{banfield1993model}, this has not been addressed in network analysis. Through clustering edges and analyzing their weights, we gain a distinct advantage in identifying ``noisy edges''. The notion of a ``noisy edge'' refers to edges that do not conform to the patterns or structures present in the majority of the edges in the network. For instance, in the email network, while a high-level manager may send regular emails to various teams under her supervision (which may share the same nodes between them), she may send a spurious email to everyone regarding a phishing attempt, or she may simply relay a message on behalf of someone else.  The first example email highlights meaningful structure in a communication network, whereas the latter two may only serve to obscure the team structure. In a contact tracing network, this could be contacts caused by pedestrians passing by each other. The presence of noisy edges can sometimes pose challenges in clustering analysis, as they may obscure the underlying structure of the network, lead to over- or underestimation of the number of clusters, or lead to misinterpretations of the clusters. Therefore, it is essential to account for noisy edges and handle them appropriately in clustering algorithms to ensure accurate and meaningful results. Though such noisy edges widely exist in real-world networks, to the authors’ knowledge, currently no algorithms have been proposed to identify them in networks.

In this paper, we propose the Weighted  Edge Clustering Adjusting for Noise (WECAN) model to address these two methodological research gaps by clustering weighted directed edges and identifying the noisy edges. The WECAN model automatically selects the number of clusters, makes use of edge weights to help inform clustering patterns, and incorporates a noise component to accound for edges which act to obscure the underlying clustering structure of the network.  WECAN is flexible in the types of edge weights that can be modeled by using a general exponential family modeling approach. Estimation is achieved by a variational Bayes generalized EM (VB-GEM) algorithm.

The remainder of the paper is as follows.  Section \ref{sec:methods} describes the proposed edge clustering model and the VB-GEM estimation algorithm.  Section \ref{sec:simstudy} describes a simulation study in which we compare our proposed approach to the existing state-of-the-art.  Section \ref{sec:pattransfers} illustrates our proposed approach by applying WECAN to patient transfer networks.  Finally, Section \ref{sec:conclusion} provides a brief discussion.

\section{Models and Methodology}
\label{sec:methods}
\subsection{Setup and notation}

Consider the case where we observe a directed weighted network with $n$ nodes represented by the set $\~A$ and $M$ edges denoted by $\~E \subset \~A \times \~A$. Without loss of generality, we will assume $\~A = \{1, 2, \cdots, n\}$.The set of edges is represented as $\~E = \{ \*e_m \}^M_{m=1}$, where each edge is denoted by $\*e_m$. Each edge $\*e_m$ is a 3-dimensional vector comprised of the following: the sending node, denoted by $e_{m1}$; the receiving node, denoted by $e_{m2}$; and the weight of the edge, denoted by $w_m$. Thus, $\*e_m = (e_{m1}, e_{m2}, w_m)$. It is important to note that self-loops are not considered in our model. For clarity, we illustrate this with a running example from the domain of hospital epidemiology, namely patient transfer networks where nodes signify hospitals, and edges signify patient transfers between hospitals. Here, $e_{m1}$ represents the sending hospital, $e_{m2}$ represents the receiving hospital and $w_m$ represents the number of patients transferred between these hospitals.

We presuppose the existence of a maximum of $(K + 1)$ latent clusters for edges in this network, comprising up to $K$ genuine clusters and 1 noise cluster. Edges within the ``noise cluster'' do not adhere to any distinct cluster structure, while edges within a ``genuine cluster'' exhibit meaningful patterns or structures within the network. Each edge is assigned to a specific latent cluster, represented by the binary variable $Z_{mk}$, where $Z_{mk} = 1$ if edge $\*e_m$ belongs to cluster $k$, and $Z_{mk} = 0$ otherwise. In the patient transfer network context, these edge clusters can be understood as groups of patient transfers exhibiting distinct patterns. For instance, transfers within a cluster may involve patients who are moving ``in network'' based on insurance coverage, moving based on specialty clinics, or moving within a certain geographical region. Meanwhile, the noisy cluster could encompass rare or unexpected transfers, such as patients having health emergencies while traveling.

In the context of clustering nodes, it is well known that various topological features may obfuscate the clustering structure of the network \citep[hence, for example, the development of the degree-corrected stochastic blockmodel proposed by][]{PhysRevE.83.016107}.  To avoid this from occurring in edge clustering, we utilize the following node-specific attributes:

\begin{itemize}
    \item $\boldsymbol{S_1} = (S_{11}, S_{12}, \cdots, S_{1n})$: representing overall propensities of nodes to send a large or small number of edges.
    \item $\boldsymbol{R_1} = (R_{11}, R_{12}, \cdots, R_{1n})$: representing overall propensities of nodes to receive a large or small number of edges.
    \item $\boldsymbol{S_2} = (S_{21}, S_{22}, \cdots, S_{2n})$: representing overall propensities of nodes to send edges with large or small weights.
    \item $\boldsymbol{R_2} = (R_{21}, R_{22}, \cdots, R_{2n})$: representing overall propensities of nodes to receive edges with large or small weights.
\end{itemize}

These attributes address degree heterogeneity as well as heterogeneity in expected edge weights.  In the patient transfer example, a large academic university will typically send and receive a large number of edges, and such a university that transfers patients from and to large urban hospitals will typically have edges with larger weights than one transferring patients from and to smaller rural hospitals.

Beyond degree heterogeneity and heterogeneity in expected edge weights, a core feature found uniquitously in networks is that of homophily, the phenomenon where nodes that are close in some feature space have a higher probability of having an edge between them.  We estimate homophilic effects by representing each node in a latent $p-$dimensional space. Stacking these $n$ latent feature vectors forms an $n \times p$ matrix $\boldsymbol{U}$ for latent sending features and an $n \times p$ matrix $\boldsymbol{V}$ for latent receiving features. Originally motivated by the Aldous-Hoover theorem, this builds off of the rich class of latent space network models first put forth by \citet{hoff2002latent}.  Furthermore, we posit that the contexts in which edges form, that is, the edge clusters, can also be described using this latent feature space, and that each edge context dictates how node features interact to build stronger or weaker edges.  Notationally, we have the following:
\begin{itemize}
    \item $\boldsymbol{Y}$: a $K \times p$ matrix where the $k^{th}$ row, $\boldsymbol{Y}_k$, represents the latent features of the $k^{th}$ edge cluster, elucidating how node features interact with edge clusters to form edges.
    \item $\boldsymbol{\Lambda}$: a $K \times p \times p$ array where the $k^{th}$ $p \times p$ diagonal matrix describes how node features interact with the $k^{th}$ edge cluster to yield large or small edge weights.
\end{itemize}
 In the patient transfer network, these variables unveil the underlying characteristics shared by hospitals and patients influencing transfers, such as geographical proximity or hospital specialization in treating specific diseases.

\subsection{Model}
The Weighted Edge Clustering Adjusting for Noise (WECAN) model is developed from the latent space edge cluster (LSEC) model \citep{sewell2021model} and automated latent space edge cluster (aLSEC) model \citep{pham2024automated}. Suppose there are $(K+1)$ latent edge clusters in the observed network; $Z_{m0} = 1$ represents that the edge is in the noise cluster, while $Z_{mk} = 1$, $k > 0$, represents that the edge is belonging to the cluster $k$. The WECAN model is given by: 

\begin{align}
    \nonumber
    \pi(\~E | Z)  
    & = 
    \prod^M_{m=1} \prod^K_{k=0} (\pi(\*e_m|Z_{mk}=1))^{Z_{mk}} 
    & \\ 
    & = 
    \prod^M_{m=1} \prod^K_{k=0} \Big(\pi(e_{m_1}|Z_{mk}=1)\times \pi(e_{m_2}|e_{m_1}, Z_{mk} = 1) \times \pi(w_m|e_{m_2},e_{m_1}, Z_{mk} = 1) \Big)^{Z_{mk}}
    &
    \label{eq:WECANmodel}
\end{align}

For the first two components in \eqref{eq:WECANmodel}, we assume that for a noisy edge, the sending and receiving nodes should be completely random and, therefore, follow a uniform distribution. For a non-noisy, or structural, edge, we mirror the LSEC and aLSEC models, letting the sending and receiving nodes involved in the $m^{th}$ edge depend on the nodes' overall propensities $(\boldsymbol{S_1}, \boldsymbol{R_1})$, the nodes' latent features $(\boldsymbol{U}, \boldsymbol{V})$, and the interactions between nodes' features and the edge's cluster $(\boldsymbol{UY}, \boldsymbol{VY})$. This leads to the following:

\begin{equation} \nonumber
     \pi(e_{m1} = i|Z_{mk}=1) =\left\{
\begin{array}{cll}
\frac{1}{N}       &      &  {k = 0}\\ \\
 \frac{e^{S_{1i}+\*U_i\*Y^T_k}}{f_{uk}}     &      & {o.w.,}
\end{array} \right.
\end{equation}

\begin{equation} \nonumber
    \pi(e_{m2} = j |e_{m_1} = i, Z_{mk}=1) =\left\{
    \begin{array}{cll}
\frac{1}{N - 1}       &      &  {k = 0 , i \neq j}\\ \\
\frac{e^{R_{1j}+\*V_j\*Y^T_k}}{f_{vk} - e^{R_{1i} + \*V_i\*Y_k^T}}     &      & {k \neq 0, i \neq j} \\ \\
0     &      & {i = j},
\end{array} \right.
\end{equation}
where: $f_{uk} = \sum ^n_{i=1}e^{S_{1i}+\*U_i\*Y^T_k}$, $ f_{vk} = \sum ^n_{i=1} e^{R_{1i} + \*V_i\*Y^T_k} $. The quantities $f_{uk}$ and $f_{vk}$ are introduced for normalization. 

We assume that the weight of a noisy edge follows an exponential distribution with user-defined rate $\lambda_a$; this value should typically be large to induce a small expected weight for those edges in the noise component. The weight of a non-noisy edge follows an exponential family distribution with canonical parameter as $\eta_{ijk}$ and dispersion parameter as $\phi_k$. Therefore we are able to model both continuous and discrete edge weights.

\begin{align}
    \nonumber
    & \pi(w_m = w|e_{m_1} = i, e_{m2} = j, Z_{mk}=1) 
    &\\
    & =
\begin{cases}
\lambda_a \exp(\lambda_a w) &  {k = 0}\\ \\ 
 h(w,\phi_k)\exp\left(\frac{\eta_{ijk}w - A(\eta_{ijk})}{a(\phi_k)} \right) 
/ \Pr(w \neq 0|\eta_{ijk}) & {o.w.},
\end{cases}  &
 \label{eq:WECANmodelp3}
\end{align}

The canonical parameter $\eta_{ijk}$ is decided by weights-related parameters consisting of the sending and receiving nodes' overall propensities $(S_{2i}, R_{2j})$ to be involved in large or small edge weights, the interaction between the nodes' and edge cluster's features $U_i \Lambda V_j'$, and the cluster-specific intercept $\beta_k$. Thus we set

\begin{equation} \label{eq:exponfamily} \nonumber
    \eta_{ijk} :=  \beta_k + S_{2i} + R_{2j} +\*U_i\*\Lambda_k \*V_j^T. 
\end{equation}

The remaining elements in \eqref{eq:WECANmodelp3} are determined by the specific exponential family distribution chosen by the user. Taking the normal distribution as an example, we have
\begin{align*}
   h(w_m,\phi_k) 
   & \propto 
   \frac{1}{\phi_k}\exp(\frac{-w_m^2}{2\phi_k^2}),  
   & \\
   A(\eta_{ijk})
   & =
   \frac{\eta_{ijk}^2}{2},
   &\\
   a(\phi_k) 
   & = 
   \phi_k^2.
   &
\end{align*}
Note that  $\phi_k$ is the standard deviation in normal distribution.

As in the aLSEC model, we propose a sparse finite mixture model for $\boldsymbol{Z_m}$, this allows for automatic detection of the number of clusters. By deliberately overfitting the model with more components than necessary, the Dirichlet prior with small expected concentration parameter encourages zero weights for irrelevant components, effectively leaving some clusters empty. This approach ensures that the model adapts to the true number of clusters without requiring prior specification.

Each cluster assignment vector $\boldsymbol{Z_m}$ is drawn from a multinomial distribution with probabilities $\boldsymbol{\pi} := (\pi_0, \pi_1 , \cdots, \pi_K)$. To make sure $\sum^{K}_{k = 0}\pi_k = 1$, we introduce $\boldsymbol{t} = c(t_0, t_1 , \cdots, t_K)$, where for the noise cluster, $\pi_0 = t_0$; for non-noise cluster, $\pi_k = t_k(1-t_0)$. So we have:
\begin{align*}
    \boldsymbol{Z_m}|\pi & \overset{iid}{\sim} Multinomial(1, \boldsymbol{\pi}) &\\
    t_0 & \sim Beta(c_0,d_0) &\\
    (t_1,\ldots,t_K) & \sim Dir(\alpha \boldsymbol{\mathds{1}_K}) &\\
    \alpha & \sim \Gamma(a_{\alpha}, b_{\alpha}) &
\end{align*}

The complete likelihood is:
\small
\begin{align} 
    \nonumber
   &
   f( \~E, \boldsymbol{Z}|\boldsymbol{\theta}) 
   & \\ \nonumber
   & = 
   \prod^M_{m=1} \left[t_0 \times \lambda_a \times \frac{1}{N(N-1)} \times exp(-\lambda_a w_m) \right]^{Z_{m0}}
   & \\ 
   &  \times  \prod^K_{k=1} \Bigg[ t_k(1-t_0) \times  
   \frac{h(w,\phi_k)\exp(\frac{\eta_{em}w_m-A(\eta_{em})}{a(\phi_k)})\times \exp(S_{1e_{m_1}} + R_{1e_{m_2}} +  \*U_{e_{m_1}}\*Y'_k + \*V_{e_{m_2}}\*Y_k')}{f_{uk}(f_{vk} - \exp(R_{1e_{m_1}} + \*V_{e_{m_1}}\*Y_k'))\Pr(w_m\neq 0|\eta_{em})} \Bigg]^{Z_{mk}} 
   &
\end{align}

\begin{figure}[H]
  \centering
  \includegraphics[scale=0.5]{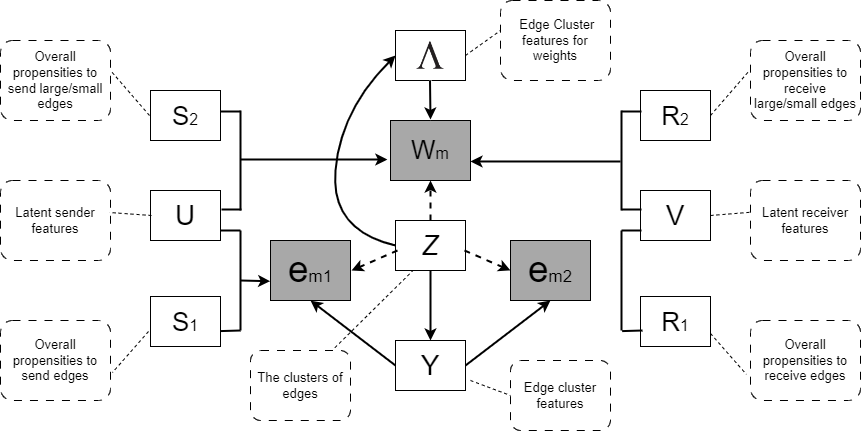}
  \caption{Schematic of the WECAN model.}
    \label{fig:schematic}
\end{figure}

Figure \ref{fig:schematic} shows all the parameters for the WECAN model. Note that if we ignore the weights of the network, WECAN model will be equivalent to LSEC model.

\subsection{Parameter Estimation}
We propose performing Bayesian estimation in the same manner as in \citet{pham2024automated}. For unknown quantities $ (\boldsymbol{S_1}, \boldsymbol{S_2}, \boldsymbol{R_1}, \boldsymbol{R_2}, \boldsymbol{U}, \boldsymbol{V}, \boldsymbol{\Lambda}, \boldsymbol{Y}, \boldsymbol{\beta}, \phi) $, following priors are assumed:

\vbox{
\begin{align} \nonumber
(\*U_i, \*V_i)^T &\overset{iid}{\sim} MVN(\*0, \boldsymbol{\Sigma_{UV}} \otimes \boldsymbol{I_p} ), &  \boldsymbol{\Sigma_{UV}} &\sim IW(\boldsymbol{\Psi_{0UV}}, \upsilon_{0UV}), 
\\ \nonumber
(S_{1i}, R_{1i})^T &\sim MVN(\*0, \boldsymbol{\Sigma_{SR1}}), &   \boldsymbol{\Sigma_{SR1}} &\sim IW(\boldsymbol{\Psi_{0SR1}}, \upsilon_{0SR1}),
\\ \nonumber
(S_{2i}, R_{2i})^T & \sim MVN(\*0, \boldsymbol{\Sigma_{SR2}}), &   \boldsymbol{\Sigma_{SR2}} & \sim IW(\boldsymbol{\Psi_{0SR2}}, \upsilon_{0SR2}), 
\\ \nonumber
\boldsymbol{Y_k} & \sim MVN(\boldsymbol{0},\boldsymbol{I_p}), &   diag(\boldsymbol{\Lambda_k}) & \sim MVN(\boldsymbol{0}, \lambda \boldsymbol{I_p}), 
 \\ \nonumber
 \boldsymbol{Z_m} &\sim Multinomial(1, \boldsymbol{\alpha}), & \boldsymbol{\alpha} & \sim Dir(\alpha_0 \boldsymbol{\mathds{1}_K)},  
 \\ \nonumber
 \beta_k \ &\propto \ 1, & \lambda & \sim \Gamma^{-1}(a_0,b_0), &
\\ \nonumber
\end{align}
}
\noindent where $MVN(\boldsymbol{a}, \boldsymbol{\Sigma})$ represents the multivariate normal distribution with $\boldsymbol{a}$ as the mean vector and $\boldsymbol{\Sigma}$ as the covariance matrix, $IW(\boldsymbol{\Psi}, \upsilon)$ represents the Inverse Wishart distribution with $\boldsymbol{\Psi}$ as scale matrix, $\upsilon$ as degrees of freedom, $Dir(\boldsymbol{a})$ denotes the Dirichlet distribution with $\boldsymbol{a}$ as concentration parameter, $\Gamma^{-1}(a, b)$ denotes inverse Gamma distribution with $a$ as shape parameter and $b$ as scale parameter. $\boldsymbol{I_p}$ is $p \times p$ identity matrix, $\boldsymbol{\mathds{1}_K}$ is the K-dimensional vector of ones. The prior on the dispersion parameter will be dependent on the specific exponential family used for the edge weights.  For example, in Section \ref{sec:pattransfers} we assume a normal distribution where we use a half-$t$ prior with $\nu_0$ denoting the degrees of freedom and $\eta_0$ denoting the scale parameter on the standard deviation:
$$
\phi_k \sim \mbox{half-}t(\nu_0,\eta_0) . 
$$

Then we propose a variational-Bayes generalized expectation–maximization (VB-GEM) algorithm \citep{dempster1977maximum,bernardo2003variational}. In the E-step, both $\mathbf{Z}$ and $\boldsymbol{t} = {t_0, t_1, \cdots, t_k}$ are the latent variables. In the M-step, we update the unknown parameters $\boldsymbol{\theta} := \{\boldsymbol{S_1}, \boldsymbol{S_2}, \boldsymbol{R_1}, \boldsymbol{R_2}, \boldsymbol{U}, \boldsymbol{V}, \boldsymbol{\Lambda}, \boldsymbol{Y}, \boldsymbol{\Sigma_{SR_1}}, \boldsymbol{\Sigma_{SR_2}}, \boldsymbol{\Sigma_{UV}}, \boldsymbol{\lambda}, \boldsymbol{\alpha} \}$ by performing a conjugate gradient approach that maximizes the expected log posterior $Q(\boldsymbol{\theta}|\boldsymbol{\theta^{t}})$, where $\boldsymbol{\theta^{t}}$ denotes the current estimates of the parameters.  

For the expectation step (E step), the variational distributions are selected by maximizing evidence lower bound (ELBO) of the log marginal likelihood, which can be derived as below:
$$ 
ELBO := E_{q(\boldsymbol{Z}, \boldsymbol{t})}\log(f(\boldsymbol{\theta}, \boldsymbol{Z}, \boldsymbol{t}|\boldsymbol{\~E})) - E_{q(\boldsymbol{Z}, \boldsymbol{t})}\log(q(\boldsymbol{Z}, \boldsymbol{t})). 
$$

The full log posterior and the details of ELBO can be found in the \ref{sec:appendixA}. In short, the full conditional of $\boldsymbol{Z}$ is the multinomial distribution; the full conditional of $\boldsymbol{t}$ is a Dirichlet distribution; and the full conditional of $t_0$ is a Beta distribution. This semi-conjugacy lends itself to a mean-field variational Bayes estimation of the means for the E step of the algorithm, providing us closed form solutions when iteratively updating the parameters of the variational approximation distribution for $\boldsymbol{Z}$, $t_0$, and $\boldsymbol{t}$ until convergence.

For the maximization step (M step), we propose a coordinate ascent approach similar to Sewell (2021) to find: 
$$ 
\boldsymbol{\theta^{t+1}} = \mbox{argmax}_{\boldsymbol{\theta}}(Q(\boldsymbol{\theta}|\boldsymbol{\theta^{t}})) 
$$
The updates of the $\{\boldsymbol{S_1}, \boldsymbol{S_2}, \boldsymbol{R_1}, \boldsymbol{R_2}, \boldsymbol{U}, \boldsymbol{V}, \boldsymbol{\Lambda}, \boldsymbol{Y} \}$ are by conjugate gradient; the updates of the $\{\boldsymbol{\Sigma_{SR_1}}, \boldsymbol{\Sigma_{SR_2}}, \boldsymbol{\Sigma_{UV}}, \boldsymbol{\lambda}, \boldsymbol{\alpha} \}$ are from analytical solutions. The details of gradients can be found in \ref{sec:appendixB}, and the analytical solutions can be found in \ref{sec:appendixC}.

To obtain a more reasonable initialization, we first run the aLSEC model once to get initialization for the variables $(\boldsymbol{S_1}, \boldsymbol{R_1}, \boldsymbol{U}, \boldsymbol{V}, \boldsymbol{Y})$; for the initial values of $\boldsymbol{S_2}$ and $\boldsymbol{R_2}$, the sum of in-degree and out-degree edge weights of the adjacent edges for each node was calculated; and when the weights were fit using a normal distribution, the standard deviation of the weights was used to set the initial value of $\phi$. The convergence in the E-step was evaluated by monitoring the changes in the ELBO. Convergence of the whole algorithm was based on a lack of change in the cluster assignments between iterations.

In the analyses of Sections \ref{sec:simstudy} and \ref{sec:pattransfers}, we repeated the VB-GEM estimation procedure 15 times with different random starting points and computed the integrated completed likelihood (ICL) to select the optimal (highest) result \citep{biernacki2000assessing}.

\section{Simulation Studies}
\label{sec:simstudy}
\subsection{Study Design}

We simulated networks with edges in 5 clusters: 1 noise cluster and 4 non-noise clusters. We varied the proportion of noisy edges in the set $\{ 0\%, 5\%, 10\%, 15\%, 20\% \}$. Under each noisy edge proportion setting, we simulated 200 networks, each with 400 nodes and an average of 7065 edges. Each pair of $(\boldsymbol{S_1}, \boldsymbol{R_1})$ and $(\boldsymbol{S_2}, \boldsymbol{R_2})$ are sampled from two independent bivariate normal distributions with mean 0, variances 2, and correlation 0.75. We set $p = 2$, and the row vectors of the latent features $\boldsymbol{Y}$ were equally spaced on a circle. The direction of the $\*U_i$’s and $\*V_i$’s were drawn from a mixture of von Mises–Fisher distributions with concentration parameter of 50, and their magnitudes were drawn from a gamma distribution with shape equal to 150 and rate equal to 40. The $\beta_k$ were simulated by a random sample without replacement from $\{-1, -2, 1, 2\}$, and $\boldsymbol{\Lambda_k}$ was set to be a $2 \times 2$ diagonal matrix with the diagonal as $\{0.4, 0.4\}$, $\{0.4, -0.4\}$,  $\{-0.4, 0.4\}$, $\{-0.4, -0.4\}$ for each non-noisy cluster respectively. The weights of edges in the noise cluster follow a gamma distribution with shape equal to 2 and rate equal to 20. The edge weights within the genuine clusters are modeled as following a normal distribution with a mean of $\beta_k + S_{2i} + R_{2j} + \*U_i\*\Lambda_k\*V_j^T$, while the standard deviations are sampled from a uniform distribution, $Unif(0.05, 0.5)$.

\begin{figure}[H]
\captionsetup[subfigure]{labelformat=empty}
      \centering
        \begin{subfigure}{0.52\linewidth}
		 \includegraphics[width=\linewidth]{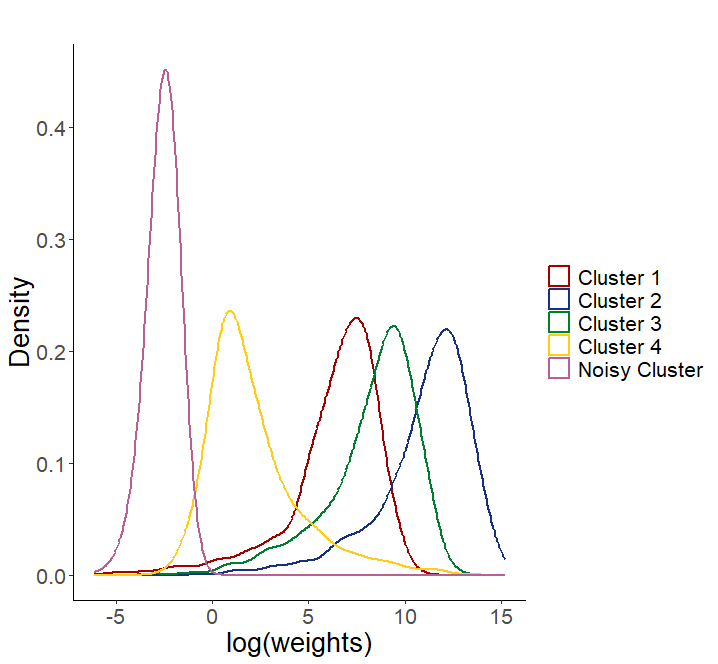}
	      \end{subfigure}
       \begin{subfigure}{0.4\linewidth}
		 \includegraphics[width=\linewidth]{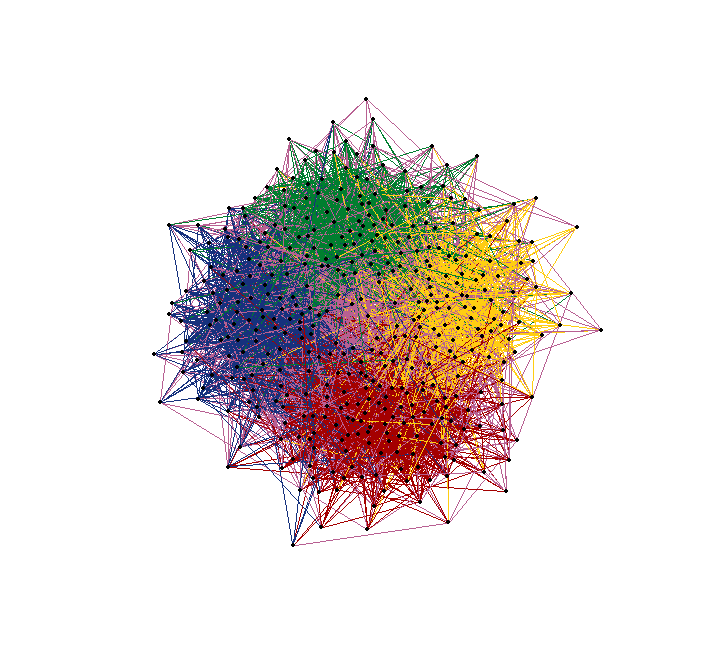}
	      \end{subfigure}
	\caption{Example of simulated network.  The yellow, red, green, and blue clusters represent meaningful network structures with varying but overlapping edge weights, whereas the mauve cluster is comprised of pairs selected uniformly at random with smaller edge weights.(Color online.)}
	\label{fig:simulation_design}
\end{figure}

Figure \ref{fig:simulation_design} presents an example of a simulated network, with different colors indicating the clusters to which the edges belong. The pink edges represent noisy edges, uniformly distributed across the network. In the left plot, the densities of log-transformed edge weights are shown, helping to highlight the distinction between noisy and non-noisy edges. While the noisy edges generally have smaller weights, the weights of non-noisy edges are heavily overlapping. The right plot displays the same network structure without showing the weights, using the Fruchterman-Reingold layout. This visualization illustrates how noisy edges can obscure the underlying clustering structure.

For each simulated network, we clustered the edges using WECAN and two other competitors and compared their performance.  First, we used spectral clustering on the line graph (SCLG). Spectral clustering detects communities by leveraging the eigenvalues of the graph Laplacian matrix. The process involves constructing a similarity graph, computing the Laplacian matrix, and embedding the graph into a lower-dimensional space based on the smallest eigenvalues. Traditional clustering algorithms like k-means are then applied to identify clusters \citep{von2007tutorial}. In the SCLG method, we first convert the original network into a line graph, and then apply spectral clustering to the resulting line graph. Second, we used the aLSEC model.

Following \citep{pham2024automated}, we set the parameter $p$ to 4 for both the aLSEC and WECAN models to effectively capture latent space information. For the WECAN model, the non-noisy edges' weights are assumed to follow a normal distribution. To evaluate estimation accuracy, we compared true cluster assignments with estimated cluster assignments using the normalized mutual information (NMI) metric \citep{danon2005comparing}. NMI is a criteria that measures the similarity between clusters of the same data set; it ranges from 0 to 1, with higher values indicating better agreement between clusters. Notably, we also attempted to compare our results with those from the linkcomm R package \citep{linkcomm2011}, but due to consistently poor performance, as evidenced by NMI values approaching zero in all cases, we have omitted these results from our analysis.

We also conducted a study to evaluate the computational time of the WECAN model. Previous research has demonstrated that the aLSEC model is highly efficient, reducing run time by 10 to over 100 times compared to the original LSEC model. In our study, we compared the running times of the WECAN model and the aLSEC model by simulating 100 networks of varying sizes, ranging from 100 to 1000 nodes. Both models were applied to each of these networks for comparison. Computation for the simulation study was done on University of Iowa High-performance Computing (HPC) system. The tests of model computation time were performed on a server with an Intel(R) Xeon(R) Gold 6230 CPU @ 2.10 GHz. All code was executed by R in version 4.0.5 \citep{Ritself2023}. Package RcppArmadillo (version 0.12.6.4.0) \citep{RcppArmadillo2014} was used to increase the processing speed.

\subsection{Results}

Table 1 presents the results of applying the WECAN model to 200 simulated networks under each scenario, showing the proportion of cases in which the correct number of non-noisy clusters (K = 4) was detected. Overall, the WECAN model performed exceptionally well in identifying the correct number of clusters. In all scenarios, it successfully detected the correct number of clusters in over 97\% of the simulations. Notably, when the proportion of noisy clusters exceeded 10\%, the model accurately identified the correct number of clusters in 100\% of the simulations. 

\begin{table}[H]
\begin{tabular}{llllll}
\hline
Proportion of Noisy Edges & 0\%   & 5\%    & 10\%  & 15\%    & 20\%                      \\ \hline
\begin{tabular}[c]{@{}l@{}}Proportion of Correctly \\ Estimated \# of clusters\end{tabular} & \multicolumn{1}{c}{97.9\%} & \multicolumn{1}{c}{97.0\%} & \multicolumn{1}{c}{98.5\%} & \multicolumn{1}{c}{100\%} & \multicolumn{1}{c}{100\%} \\ \hline
\end{tabular}
\caption{Performance of the WECAN model in correctly identifying the true number of clusters.}
\end{table}

\begin{figure}
\captionsetup[subfigure]{labelformat=empty}
      \centering
	   \begin{subfigure}{0.23\linewidth}
		\includegraphics[width=\linewidth]{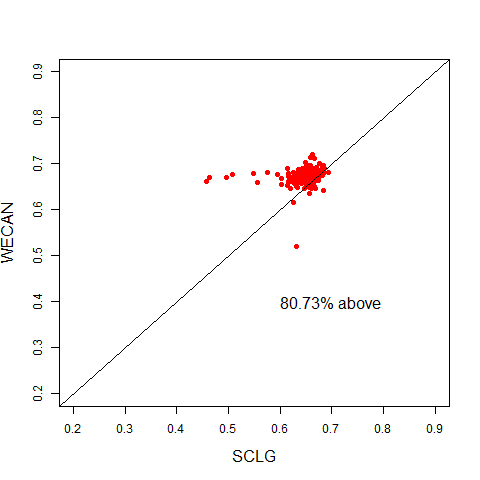}
	   \end{subfigure}
	   \begin{subfigure}{0.23\linewidth}
		\includegraphics[width=\linewidth]{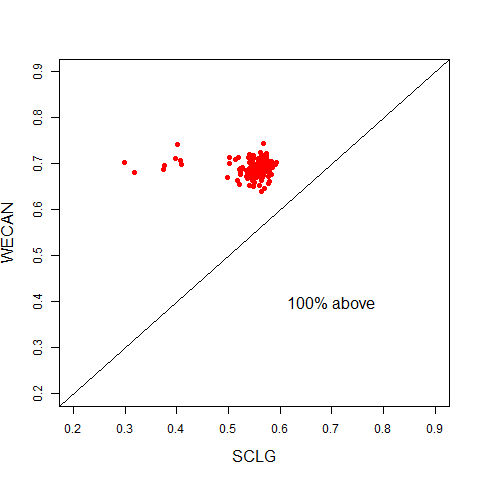}
	    \end{subfigure}
          \begin{subfigure}{0.23\linewidth}
		\includegraphics[width=\linewidth]{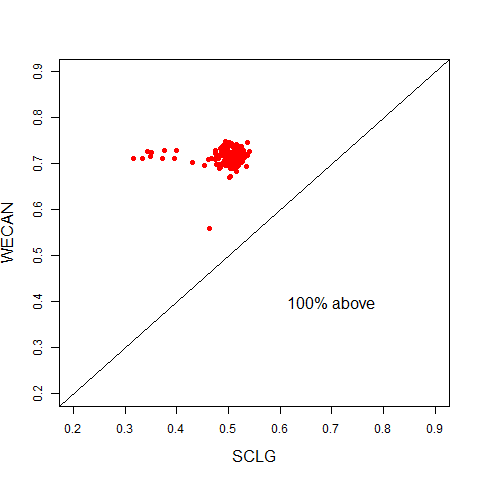}
	    \end{subfigure}
           \begin{subfigure}{0.23\linewidth}
		\includegraphics[width=\linewidth]{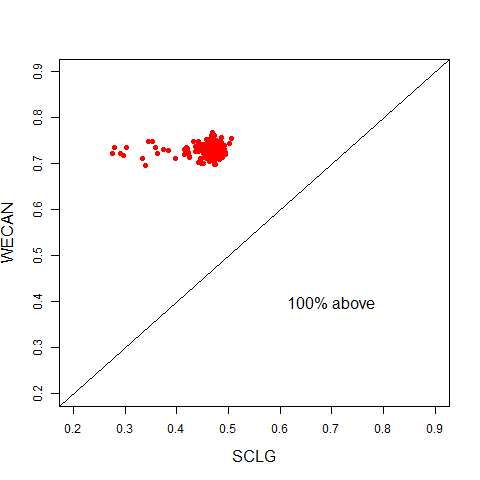}
	    \end{subfigure}
	\vfill
	     \begin{subfigure}{0.23\linewidth}
		 \includegraphics[width=\linewidth]{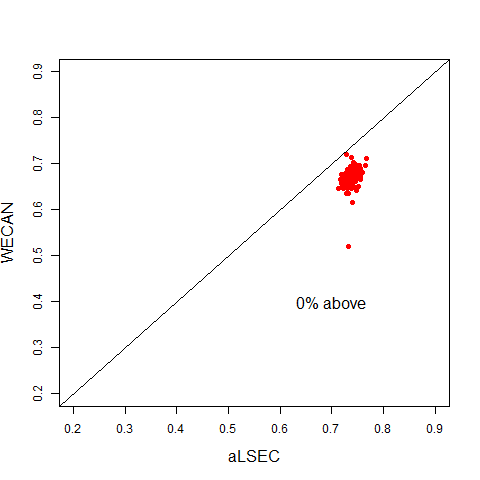}
		  \caption{0\% Noisy Edge}
	      \end{subfigure}
	       \begin{subfigure}{0.23\linewidth}
		  \includegraphics[width=\linewidth]{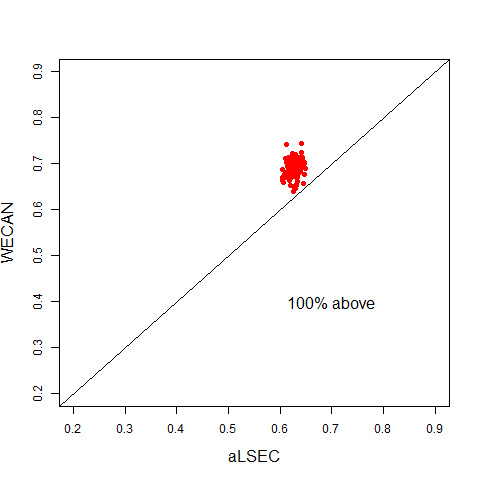}
            \caption{5\% Noisy Edges}
	       \end{subfigure}
         \begin{subfigure}{0.23\linewidth}
		 \includegraphics[width=\linewidth]{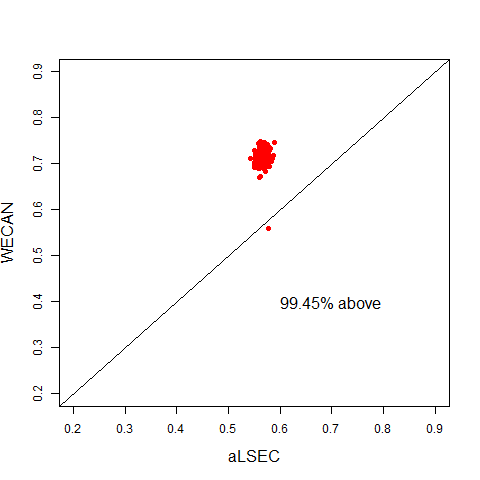}
		  \caption{10\% Noisy Edges}
	      \end{subfigure}
       \begin{subfigure}{0.23\linewidth}
		 \includegraphics[width=\linewidth]{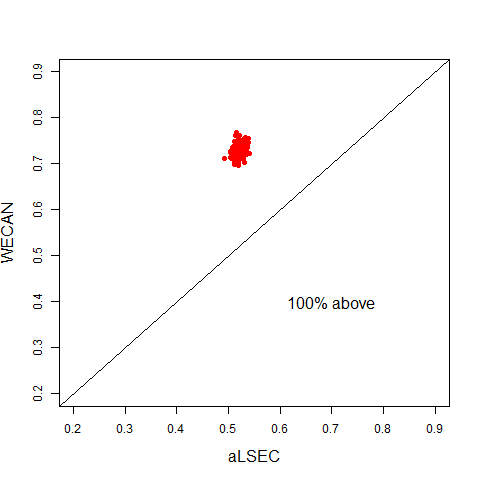}
		  \caption{15\% Noisy Edges}
	      \end{subfigure}
	\caption{NMI comparison of SCLG, aLSEC model and WECAN model in our simulation study.}
	\label{fig:simulation_results}
\end{figure}

Figure \ref{fig:simulation_results} presents a visual comparison of the Normalized Mutual Information (NMI) for the estimation results obtained from spectral clustering on the line graph, aLSEC model, and WECAN model. The top row depicts comparisons between spectral clustering on the line graph and the WECAN model, while the bottom row illustrates comparisons between the aLSEC model and the WECAN model. Each red point represents the NMI obtained from spectral clustering on the line graph or the aLSEC model plotted against the NMI from the WECAN model for a simulated network. The black line indicates the line $y = x$, where points above it indicate instances where the WECAN model outperforms the other model, while points below it denote the opposite.

The WECAN model consistently outperforms spectral clustering on the line graph across all noisy proportion settings.  Remarkably, even in scenarios where no noisy edges are present and an incorrect model is assigned to the WECAN model, it still achieves better prediction accuracy in over $80 \%$ of the networks.

When there are no noisy edges present in the simulated network, the aLSEC model demonstrates superior prediction compared to the WECAN model. However, even a small proportion of noisy edges (as low as $5\%$ ) leads to an improvement in prediction accuracy for the WECAN model for all simulations. While not shown, these results hold true when the user-input parameter $p$ is varied from 4 to 2, indicating the robustness of the WECAN model across different settings.

\begin{figure}[H]
  \centering
  \includegraphics[scale=0.35]{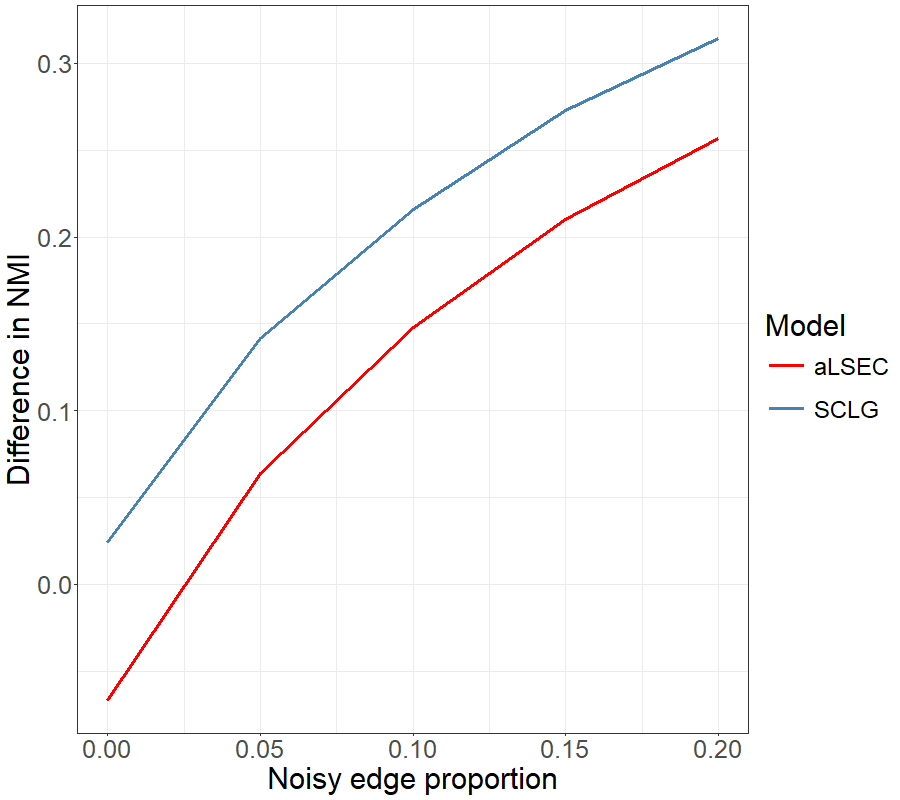}
    \caption{Mean difference in NMI across noisy proportion settings}
    \label{fig:4}
 \end{figure}

Figure \ref{fig:4} shows the mean difference in NMI across various noisy proportion settings. The blue curve represents the difference between the mean NMI of the WECAN model and SCLG, while the red curve represents the difference between the mean NMI of the WECAN model and the aLSEC model. Both differences increase as the proportion of noisy edges rises. At a noisy edge proportion of $20 \%$, the WECAN model outperforms SCLG by 0.31 in NMI and the aLSEC model by 0.26.

\begin{figure}[H]
  \centering
  \includegraphics[scale=0.42]{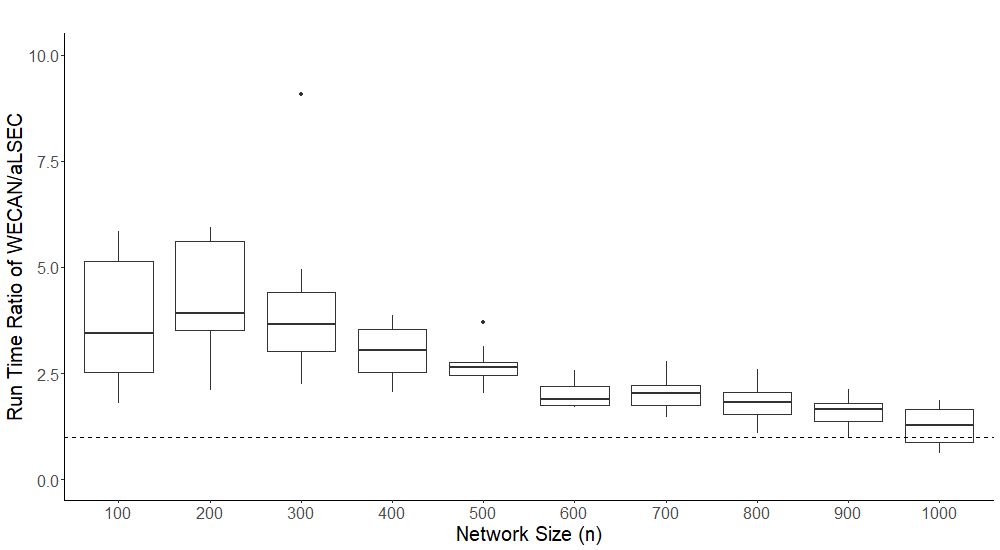}
    \caption{The ratios of the run time of the WECAN model over the aLSEC model.}
    \label{fig:5}
 \end{figure}

\begin{figure}
\captionsetup[subfigure]{labelformat=empty}
      \centering
         \begin{subfigure}{0.45\linewidth}
		 \includegraphics[width=\linewidth]{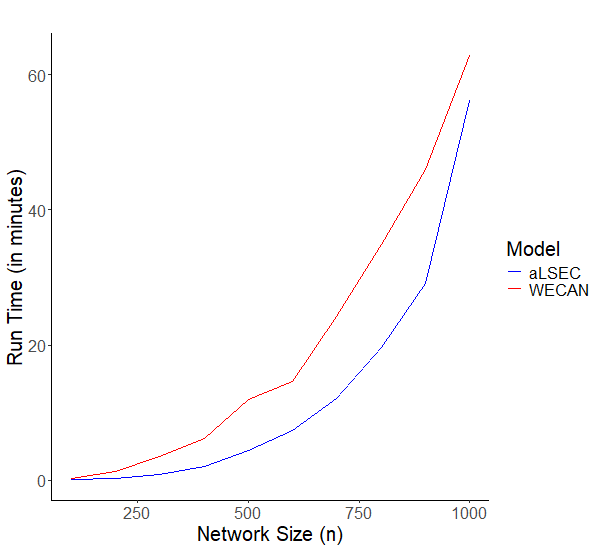}
	      \end{subfigure}
       \begin{subfigure}{0.45\linewidth}
		 \includegraphics[width=\linewidth]{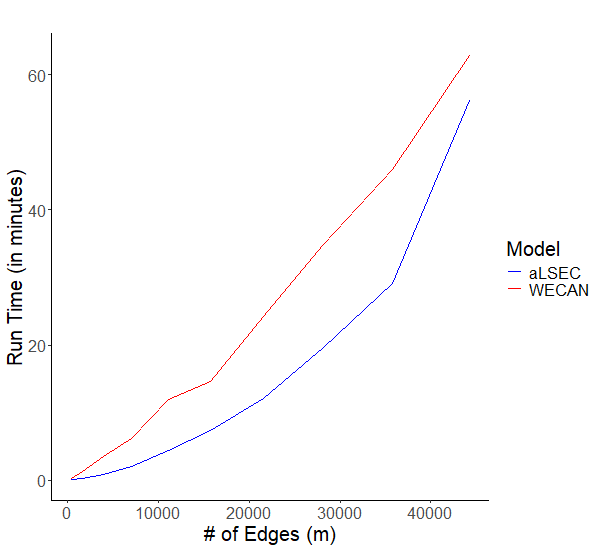}
	      \end{subfigure}
	\caption{Analysis of computational time on synthetic networks.}
	\label{fig:5-2}
\end{figure}

Figure \ref{fig:5} compares the running times of the aLSEC model and the WECAN model. The results show that while the aLSEC model generally runs faster than the WECAN model, the difference in computational time decreases as the network size increases. This is because the WECAN model uses the aLSEC model as an initial step, providing a strong starting point that reduces the additional computational time required. For smaller networks, the WECAN model takes approximately three times longer than the aLSEC model, but since the overall time is still short, this difference is less impactful. For larger networks, the running times of the two models converge, ensuring that the WECAN model maintains a reasonable computational time overall. Figure \ref{fig:5-2} presents the actual runtime (in minutes) as a function of the network's number of nodes ($n$) and edges ($m$), respectively. The WECAN model shows only a slight increase in runtime compared to the aLSEC model. With a computational complexity of ${\cal O}(N + M)$, the right plot highlights that for large and dense networks where $m \gg n$, the runtime grows approximately linearly with the number of edges in the network. 

\section{Real Data Example}
\label{sec:pattransfers}

To illustrate the effectiveness of the WECAN model on real-world data, we conducted an analysis using patient transfer network from New York state. These datasets were obtained from the Healthcare Cost and Utilization Project (HCUP) State Inpatient Databases (SID).

This patient transfer network is constructed from the New York SID from 2005 to 2016. In this network, the nodes represent New York hospitals, and the edges indicate patients who are transferred between them. Throughout the study period, we interpreted a directed edge from hospital A to hospital B as indicating patient transfers from hospital A to hospital B. The weight of each edge corresponds to the number of patients transferred between the two hospitals during the study period. The network encompasses 167 nodes and 6187 edges, with an average weight of 67.16.

Analyzing the patient transfer network is crucial because transfer patients have been found to play a significant role in infectious disease outbreaks. Studies have demonstrated that patient transfer patterns profoundly impact the spread of healthcare-associated infections (HAI), such as hospital-acquired MRSA \citep{donker2010patient}. As patients are represented by edges in the patient transfer network, utilizing edge clustering during HAI outbreaks can provide valuable insights into transmission patterns and flow.

We applied the WECAN model to the New York State patient transfer network, assuming a log-normal distribution for the non-noisy edges.  The WECAN approach relying on sparse finite mixtures automatedly selected eight non-noise clusters. We then plotted the estimated clusters both with and without the noisy edges included. Finally, we visualized the clusters based on their geographic locations to explore potential real-world implications for these clusters.

\begin{figure}[H]
  \centering
  \includegraphics[scale=0.6]{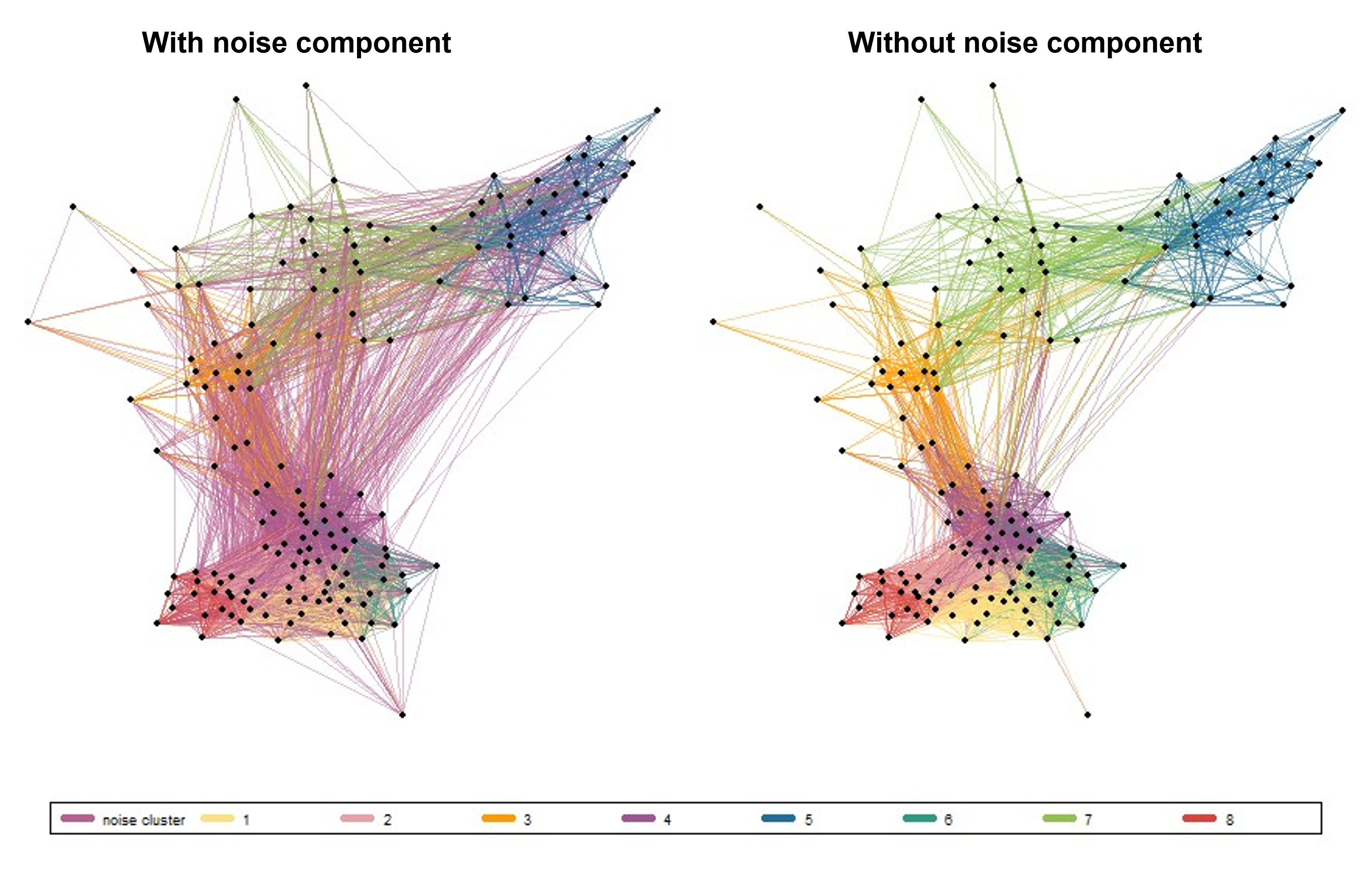}
    \caption{Clustering of the New York state patient transfer network by WECAN model.}
    \label{fig:6}
 \end{figure}

Figure \ref{fig:6} visually illustrates the edge clusters within the New York patient transfer network. Our analysis revealed 8 distinct non-noise clusters, while 1823 edges were identified as noisy edges, with an average weight of $1.1$. In the left plot of Figure \ref{fig:6}, the noisy edges are colored in pink, visually confirming our assumption that they are uniformly distributed throughout the network. Upon removing the noisy edges, as shown in the right plot, the patterns of the remaining 8 non-noise clusters become clearer. These non-noise clusters exhibit high concentration, which aligns with our intuition.

\begin{figure}[H]
\captionsetup[subfigure]{labelformat=empty}
      \centering
	   \begin{subfigure}{0.40\linewidth}
		\includegraphics[width=\linewidth]{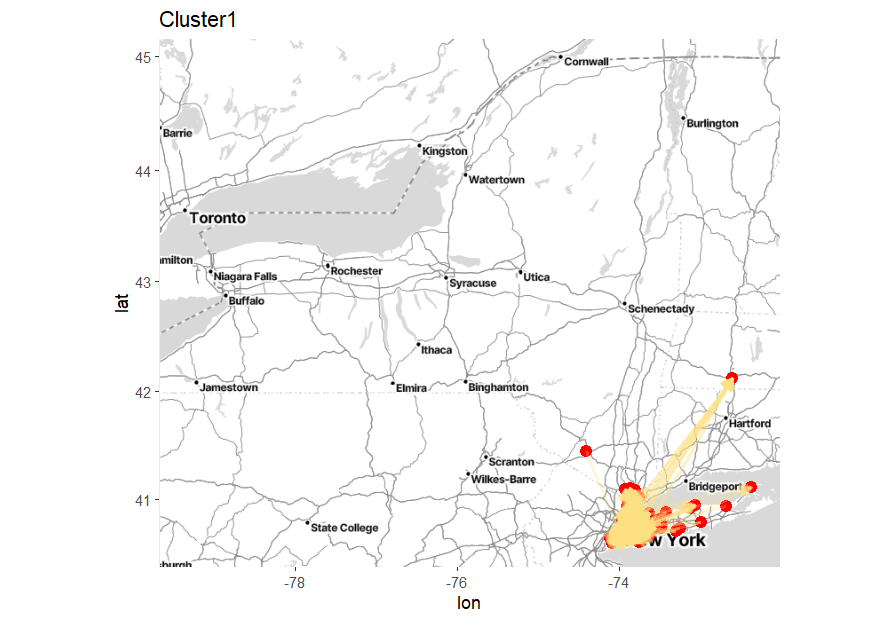}
	   \end{subfigure}
	   \begin{subfigure}{0.40\linewidth}
		\includegraphics[width=\linewidth]{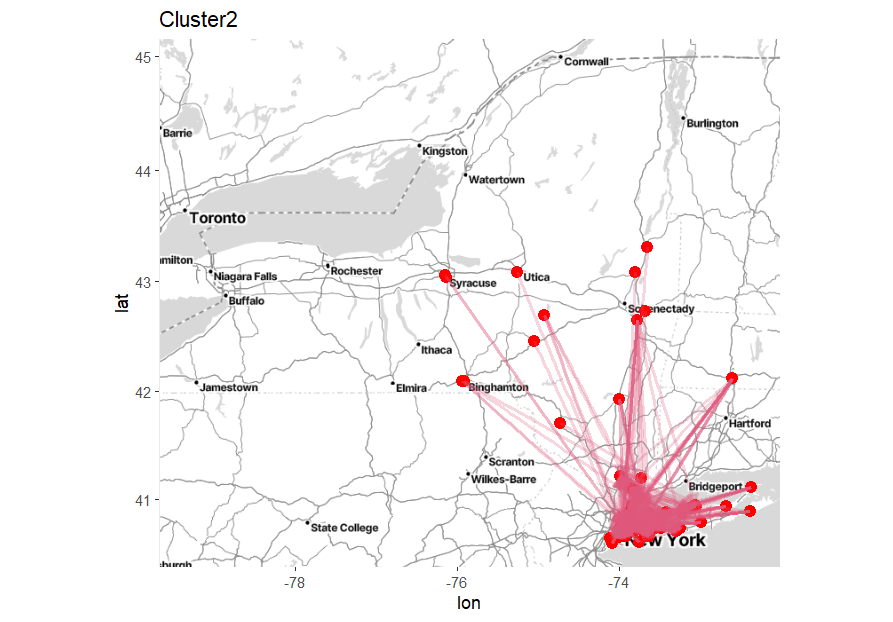}
	    \end{subfigure}
          \begin{subfigure}{0.40\linewidth}
		\includegraphics[width=\linewidth]{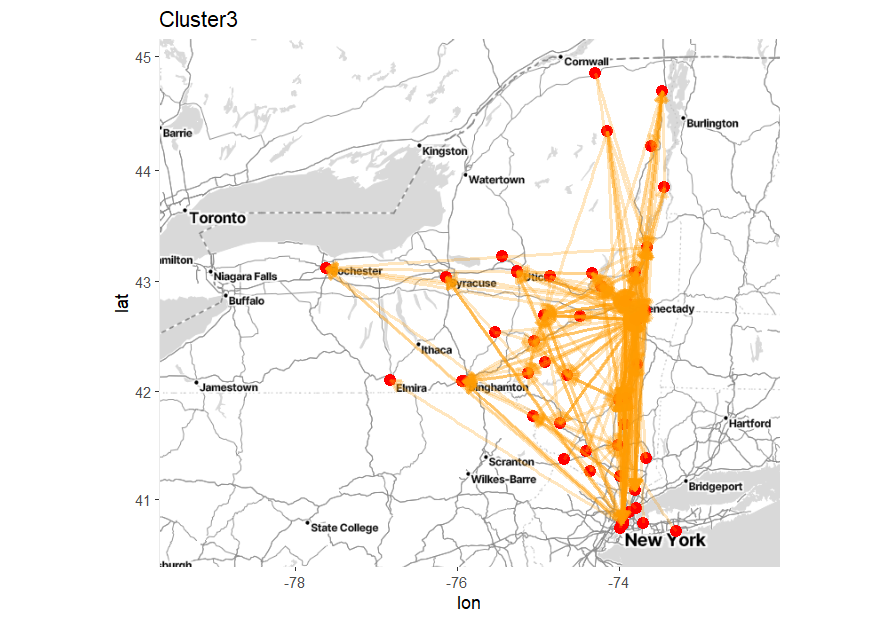}
	    \end{subfigure}
           \begin{subfigure}{0.40\linewidth}
		\includegraphics[width=\linewidth]{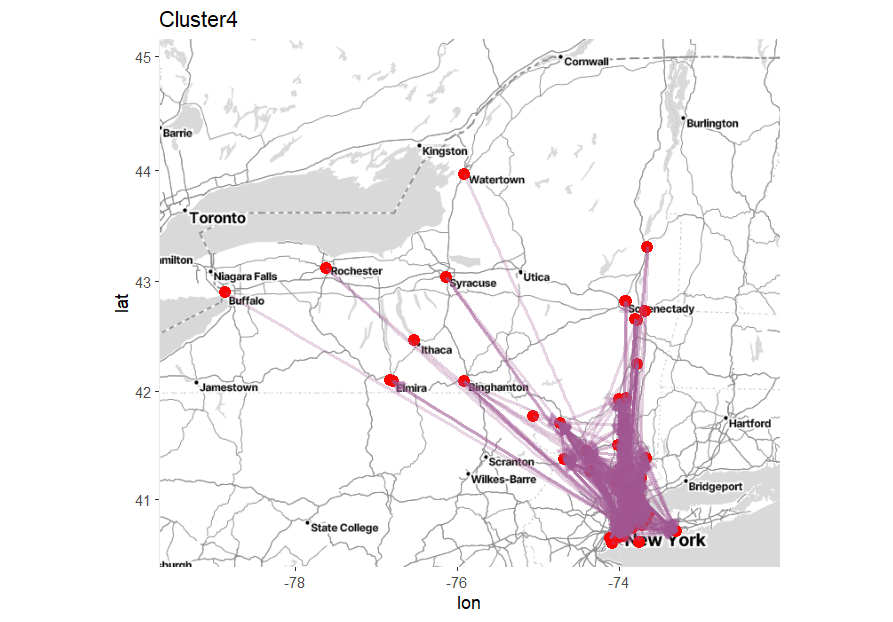}
	    \end{subfigure}
	\vfill
	     \begin{subfigure}{0.40\linewidth}
		 \includegraphics[width=\linewidth]{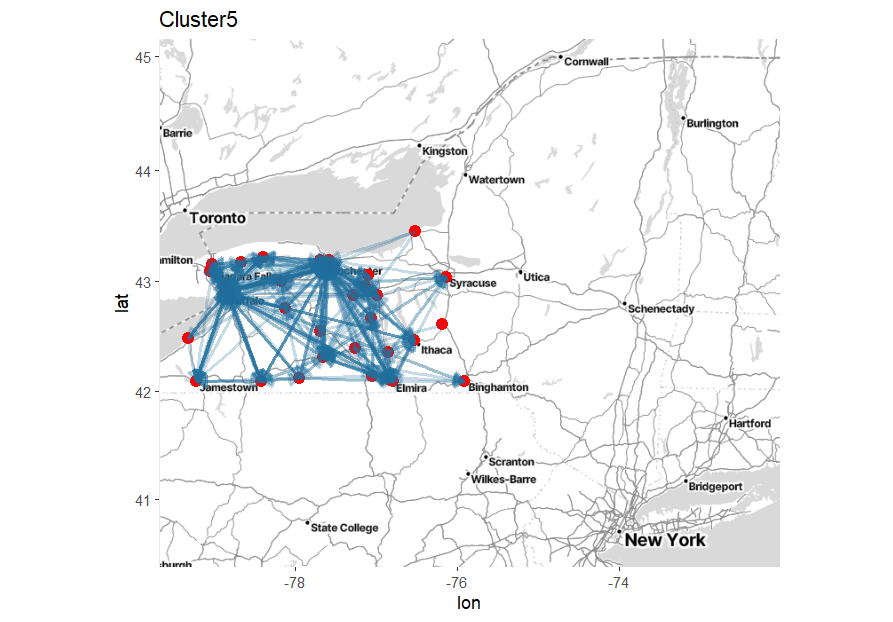}
	      \end{subfigure}
	       \begin{subfigure}{0.40\linewidth}
		  \includegraphics[width=\linewidth]{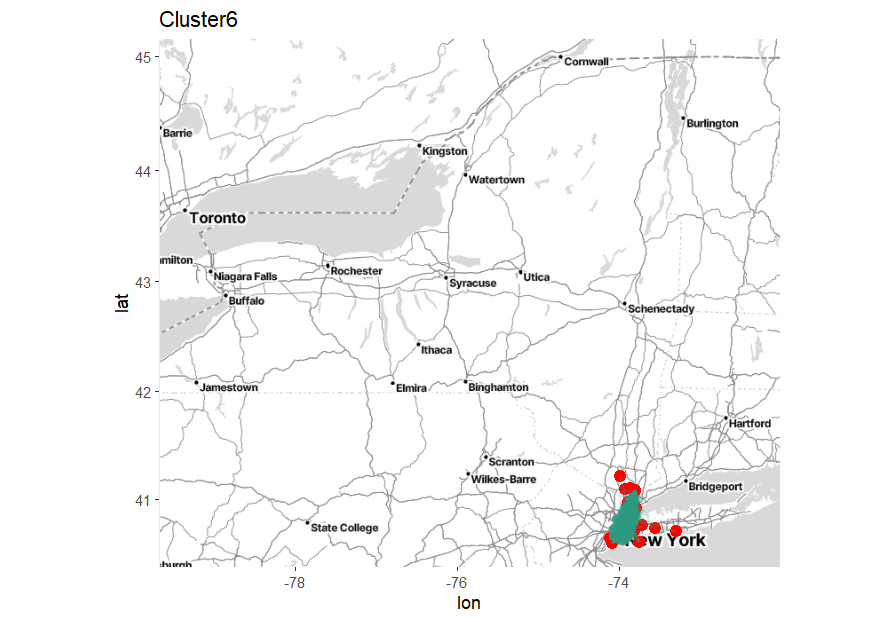}
	       \end{subfigure}
         \begin{subfigure}{0.40\linewidth}
		 \includegraphics[width=\linewidth]{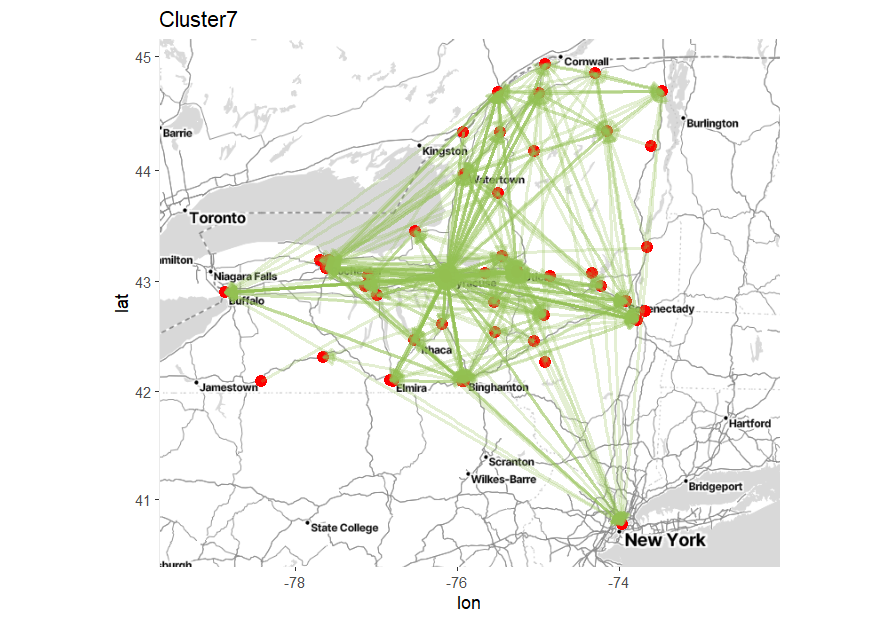}
	      \end{subfigure}
       \begin{subfigure}{0.40\linewidth}
		 \includegraphics[width=\linewidth]{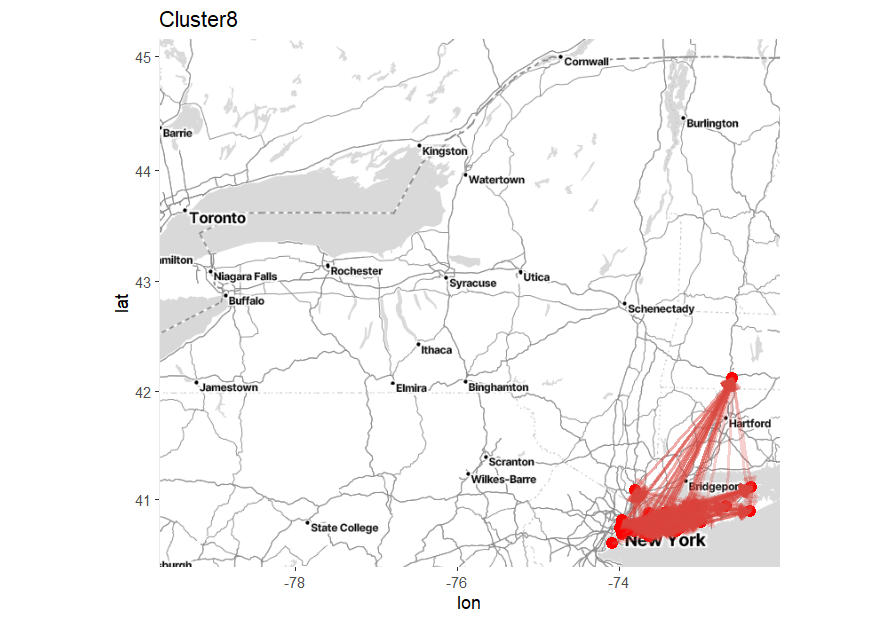}
	      \end{subfigure}
	\caption{Geographic Distribution of Non-Noisy Clusters in New York State’s Patient Transfer Network.}
	\label{fig:realworld}
\end{figure}

Figure \ref{fig:realworld} shows the geographic distribution of the eight non-noisy clusters on the map, reflecting the locations of the hospitals involved in patient transfers. This sheds light on the real-world significance of the edge clusters. For instance, Cluster 1 represents patient transfers from areas surrounding New York City into the city itself. Cluster 2 captures transfers from rural areas to New York City and Long Island. Cluster 3 corresponds to transfers directed toward the Albany Med Health System, the largest healthcare provider in the Capital District of New York. Cluster 4 represents transfers from smaller cities to New York City. Cluster 5 reflects patient transfers within the northwestern regions of New York state. Cluster 6 involves transfers within New York City itself. Cluster 7 is centered around transfers between hospitals in the Syracuse and Rochester areas. Finally, Cluster 8 represents transfers between Springfield, New York City, and Long Island. Overall, the clusters identified by the WECAN model align with meaningful real-world healthcare dynamics, indicating that the model’s results are interpretable and realistic.

We conducted a Pearson correlation test between the estimated parameters from the WECAN model and key hospital features. The results show a significant positive correlation between the estimated $\boldsymbol{S_2}$ and total admissions at the hospital (correlation = 0.26, p-value $< 0.001$). This indicates that larger hospitals, in terms of admissions, tend to have a higher propensity to send more patients. Additionally, we found a positive correlation between the hospital's estimated $\boldsymbol{R_1}$ and the total number of beds (correlation = 0.19, p-value $< 0.05$), suggesting that hospitals with more beds are more likely to receive patients. These findings are consistent with real-world expectations and demonstrate that the WECAN model produces interpretable and meaningful output parameters.

In network analysis, a common approach to mitigate the impact of such noisy edges is to remove edges with weights below a certain cutoff value. For instance, in patient transfer networks, researchers often disregard edges with a weight of 1, indicating that only one patient was transferred between two hospitals during the study period.
In the New York patient transfer network, out of the 1823 identified "noisy edges," only 914 have a weight of 1. This finding suggests that setting a cutoff value may overlook edges with weights slightly above the threshold that still mask the underlying connectivity patterns. 

We also applied the WECAN model to the Wisconsin patient transfer network, generated from SID database from 2013 to 2017. This network exhibits a much sparser structure (density $=0.10$ for WI versus $0.22$ for NY). In the Wisconsin patient transfer network example, there are 678 edges associated with a weight of 1. However, only 153 of them have been identified as "noisy edges" by the WECAN model. In this case, the WECAN model has the potential to help preserve information for network analysis by accurately identifying and excluding noisy edges. Compared with results from the NY network, these suggest that WECAN can differentiate small and unimportant edges from small and important edges.

This observation highlights the limitations of using a cutoff value alone to identify noisy edges. Therefore, a more sophisticated approach, such as the WECAN model, which takes into account the entire distribution of edge weights and other network properties, is essential for accurately identifying and removing noisy edges in network analysis.

\section{Discussion}
\label{sec:conclusion}

In this paper we have introduced the WECAN model, an innovative model-based clustering algorithm developed from the aLSEC model, incorporating the information in the edge weights and augmented with a noisy component. It has been demonstrated that dichotomizing network edges with thresholds may distort the latent structure of the network, potentially resulting in biased outcomes \citep{berardo2020collaborative}. Our model fills a significant gap in current network edge clustering methods by utilizing edge weights, thus offering a more nuanced understanding of network structures.

Furthermore, the incorporation of a noise component in the WECAN model strengthens the resilience of edge clustering in the presence of extraneous edges. Noise components are known to disrupt clustering algorithms, making it challenging to detect the cluster structure of the remaining domain points \citep{ben2014clustering}. Our real data analysis demonstrates that the WECAN model effectively distinguishes between small yet significant edges and small, less significant ones, thereby aiding in the preservation of valuable information for network analysis.

Through the simulation study, we demonstrated the effectiveness of the WECAN model in discerning meaningful patterns from complex network structures, particularly in scenarios involving noisy edges. The WECAN model can make a substantial improvement in prediction accuracy achieved by the WECAN model, underscoring its utility in practical applications.

Our analysis of patient transfer network data underscores the prevalence of unstructured ``noisy edges'' in real networks, further emphasizing the importance of robust clustering algorithms like WECAN. The ability to accurately identify and delineate edge clusters is applicable to a wide range of applications. For instance, in the context of infectious disease transmission, understanding patient clusters can facilitate proactive measures to mitigate risks and control outbreaks, while in social networks we may better understand the contexts in which relationships form.

Limitations of our proposed approach include the following.  First, we currently do not have a method for determining whether the noise component ought to be included or excluded.  Future work in selecting either a model with or without a noise component would be important for practitioners.  Second, while our method is computationally efficient for sparse networks, growing linearly with respect to the number of nodes and edges in the network, this scalability does not hold for dense networks. Third, we assumed an exponential distribution on the weights of the noise component.  This was to induce small edge weights overall while allowing some edge weights from the noise component to be large.  This may not be appropriate in all cases, e.g., when the edge weights may be negative. Despite these limitations, the WECAN model provides a novel way to uncover hidden structures and dynamics within networks, helping to improve insights and decision-making in different areas.

While the current application of the WECAN model focuses on static networks, it has the potential to be extended to longitudinal network data by incorporating time as an additional dimension. Moreover, the model could be adapted to account for node and edge attributes, providing richer insights by incorporating additional information such as node characteristics or edge-specific properties. Future work may explore these extensions to enhance the model's applicability in more complex network structures.

\section*{Acknowledgments}
This work was supported by the US Centers for Disease Control and Prevention (5 U01CK000594-04-00) as part of the MInD-Healthcare Program.

\appendix

\section{Posterior and ELBO}
\label{sec:appendixA}
The full log posterior is given by:

\begin{align} \nonumber
 & E_{q(\mathbf{Z}, \mathbf{t})}log(f(\mathbf{\theta}, \mathbf{Z}, \mathbf{t}|\~E)) 
 & \\ \nonumber
 & \propto \sum^K_{k=1} \sum^M_{m=1} E(Z_{mk}) \times \left[\log(h(w_m,\phi_k))+ \frac{\eta_{e_{m1}e_{m2}k}w_m-A(\eta_{e_{m1}e_{m2}k})}{a(\phi_k)} \right. &
 \\ \nonumber 
&\left. - \log(\Pr(w_m\neq0|\eta_{e_{m1}e_{m2}k})) + S_{1e_{m1}} + R_{1e_{m2}}+ \*U_{e_{m1}}\*Y_k' + \*V_{e_{m2}}\*Y_k' \right. &
 \\ \nonumber
&\left. - \log(f_{uk}) - \log(f_{vk} - e^{R_{1e_{m1}} + \*V_{e_{m1}}\*Y_k'}) + \log(t_k) + \log(1-t_0) \right]  &
\\ \nonumber
& + \sum^M_{m=1} E(Z_{m0}) \times (\log(t_0) - \log(\frac{N(N-1)}{\lambda_a}) - \lambda_a w_m ) & \\ \nonumber
 & + \log\Gamma(K\alpha) - K\log\Gamma(\alpha) +  (\alpha_0-1) \times \log(E(t_0)) + (\beta_0-1) \times \log(1-E(t_0))  + \sum^K_{k=1}(\alpha - 1)E(\log(t_k)) &
\\  \nonumber
& - \frac12 tr\left((S_1 \quad R_1)\*\Sigma_{SR1}^{-1} (S_1 \quad R_1)'\right) - \frac{n}{2}\log|\*\Sigma_{SR1}| &
\\ \nonumber
&- \frac12tr\left((S_2 \quad R_2)\*\Sigma_{SR2}^{-1} (S_2 \quad R_2)'\right) - \frac{n}{2}\log|\*\Sigma_{SR2}| &
\\ \nonumber
&-\frac12tr\left((\*U \quad \*V)(\*\Sigma_{UV}^{-1}\otimes \*I_p)
(\*U \quad \*V)'\right) - \frac{np}{2}\log|\*\Sigma_{UV}| &
\\ \nonumber
& + a_{\alpha}\log\alpha - b_{\alpha}\alpha - \sum_k \frac{\|\*Y_k\|^2}{2}  &
\\ \nonumber
& - \frac{\upsilon_{SR1} + p + 1}{2}\log(|\*\Sigma_{SR1}|) - \frac{1}{2}tr(\*\Phi_{SR1}\*\Sigma_{SR1}^{-1}) &
\\ \nonumber
& - \frac{\upsilon_{SR2} + p + 1}{2}\log(|\*\Sigma_{SR2}|) - \frac{1}{2}tr(\*\Phi_{SR2}\*\Sigma_{SR2}^{-1}) &
\\ \nonumber
& - \frac{\upsilon_{UV} + p + 1}{2}\log(|\*\Sigma_{UV}|) - \frac{1}{2}tr(\*\Phi_{UV}\*\Sigma_{UV}^{-1}) & 
\\ \nonumber
& -\frac{\nu_0 + 1}{2}\sum_k \log(1 + \frac{\phi_k^2}{\nu_0\eta_0^2}) -  \frac{\sum_k \|\diag(\*\Lambda_k)\|^2 + 2b_0}{2\lambda} - (a_0 + 1 + \frac{Kp}{2})\log \lambda. & 
\end{align}

Besides:
$$ E_{q(\mathbf{Z}, \mathbf{t})}log(q(\mathbf{Z}, \mathbf{t})) = \left[ \Sigma^M_{m=1}E_{q(\mathbf{Z})}(\log q(\mathbf{Z_m})) \right] + E_{q(\mathbf{t})}(\log q(\mathbf{t}))  $$

\begin{align} \nonumber
\Sigma^M_{m=1}E_{q(\mathbf{Z})}(\log q(\mathbf{Z_m})) & = E(\*Z_{m0}) \times \log\left(t_0 \times \lambda_a \times \frac{1}{N(N-1)} \times exp(-\lambda_a w_m) \right)\\ \nonumber
   &    \sum^K_{k=1} E(Z_{mk}) \times \log\Bigg(t_k(1-t_0) \times  \\ \nonumber & \frac{h(w,\phi_k)\exp(\frac{\eta_{em}w_m-A(\eta_{em})}{a(\phi_k)})\times \exp(S_{1e_{m_1}} + R_{1e_{m_2}} +  \*U_{e_{m_1}}\*Y'_k + \*V_{e_{m_2}}\*Y_k')}{f_{uk}(f_{vk} - \exp(R_{1e_{m_1}} + \*V_{e_{m_1}}\*Y_k'))\Pr(w_m\neq 0|\eta_{em})} \Bigg) \\ \nonumber
\end{align}

\begin{align} \nonumber
E_{q(\mathbf{t})}(\log q(\mathbf{t})) & = \ (E(\Sigma^M_{m=1}Z_{m0}) + \alpha_0-1) \log t_0 + (M - E(\Sigma^M_{m=1}Z_{m0}) +\beta_0-1) \log(1-t_0) \\ \nonumber
& + \Sigma^K_{k=1} (E(\Sigma^M_{m=1}Z_{mk}) + \alpha - 1) \times \log(t_k) \\ \nonumber
\end{align}

\section{The Gradients}
\label{sec:appendixB}
\begin{align} \nonumber
\frac{\partial Q}{\partial S_{1i}} & = 
\sum_k\left( p_{(i1)k} - p_{\cdot k}\frac{e^{S_{1i} + \*U_i\*Y_k'}}{f_{uk}} \right) - \Sigma_{SR1,11}S_{1i} - \*\Sigma_{SR1,12}R_{1i} &
\\ \nonumber
\frac{\partial Q}{\partial S_{2i}} & = 
\sum^K_{k=1} \sum^M_{m:e_{m1}=i}p_{mk}\left(\frac{1}{a(\phi_k)} \left(w_m - \frac{\partial A(\eta_{ie_{m2}k})}{\partial \eta_{ie_{m2}k}}\right) - \frac{\frac{\partial}{\partial \eta_{ie_{m2}k}} \Pr(w_m\neq 0|\eta_{ie_{m2}k})}{\Pr(w_m\neq 0|\eta_{ie_{m2}k})} \right) &
\\ \nonumber
&  - \Sigma_{SR2,11}S_{2i} - \Sigma_{SR2,12}R_{2i} &
\\ \nonumber
\frac{\partial Q}{\partial \*U_i} & = 
\sum^K_{k=1} \left\{\sum^M_{m:e_{m1}=i}p_{mk}\left(\frac{1}{a(\phi_k)} \left(w_m - \frac{\partial A(\eta)}{\partial \eta}\right)\*V_{e_{m2}}\*\Lambda_k - \frac{\frac{\partial}{\partial \eta_{ie_{m2}k}} \Pr(w_m\neq 0|\eta_{ie_{m2}k})}{\Pr(w_m\neq 0|\eta_{ie_{m2}k})}\*V_{e_{m2}}\*\Lambda_k \right) \right. &
\\ \nonumber
& \left. + \left(p_{(i1)k} - \frac{p_{\cdot k}}{f_{uk}}e^{S_{1i} + \*U_i\*Y_k'} \right)\*Y_k \right\}- \Sigma_{UV,11}\*U_i - \Sigma_{UV,12}\*V_i &
\\  \nonumber
\frac{\partial Q}{\partial R_{1i}} & = 
\sum_k\left( p_{(i2)k} - e^{R_{1i} + \*V_i\*Y_k'}\left(H_k - \frac{p_{(i1)k}}{f_{vk} - e^{R_{1i} + \*V_i\*Y_k'}} \right) \right) - \Sigma_{SR1,22}R_{1i} - \Sigma_{SR1,12}S_{1i} &
\\ \nonumber
\frac{\partial Q}{\partial R_{2i}} & = 
\sum^K_{k=1} \sum^M_{m:e_{m2}=i}p_{mk}\left(\frac{1}{a(\phi_k)} \left(w_m - \frac{\partial A(\eta)}{\partial \eta}\right) - \frac{\frac{\partial}{\partial \eta_{e_{m1}ik}} \Pr(w_m\neq 0|\eta_{e_{m1}ik})}{\Pr(w_m\neq 0|\eta_{e_{m1}ik})} \right) &
\\  \nonumber
&  - \Sigma_{SR2,22}R_{2i} - \Sigma_{SR2,12}S_{2i} &
\\ \label{eq:gradients}
\frac{\partial Q}{\partial \*V_i} & = 
\sum^K_{k=1}\left\{ \sum^M_{m:e_{m2}=i}p_{mk}\left(\frac{1}{a(\phi_k)} \left(w_m - \frac{\partial A(\eta)}{\partial \eta}\right)\*U_{e_{m1}}\*\Lambda_k - \frac{\frac{\partial}{\partial \eta_{e_{m1}ik}} \Pr(w_m\neq 0|\eta_{e_{m1}ik})}{\Pr(w_m\neq 0|\eta_{e_{m1}ik})}\*U_{e_{m1}}\*\Lambda_k \right) \right. &
\\ \nonumber
& \left. + \left( p_{(i2)k} - e^{R_{1i} + \*V_i\*Y_k'}\left(H_k - \frac{p_{(i1)k}}{f_{vk} - e^{R_{1i} + \*V_i\*Y_k'}} \right) \right)\*Y_k \right\} - \Sigma_{UV,22}\*V_i - \Sigma_{UV,12}\*U_i & 
\\ \nonumber
\frac{\partial Q}{\partial \diag(\*\Lambda_k)} & =
\sum_{m=1}^Mp_{mk}\left[ \frac{1}{a(\phi_k)}\left( w_m - \frac{\partial A(\eta)}{\partial \eta} \right)\big(\*U_{e_{m1}}\circ \*V_{e_{m2}} \big) \right. &
\\ \nonumber
& \left. - \frac{\frac{\partial}{\partial \eta_{e_{m1}e_{m2}k}} \Pr(w_m\neq 0|\eta_{e_{m1}e_{m2}k})}{\Pr(w_m\neq 0|\eta_{e_{m1}e_{m2}k})} \big(\*U_{e_{m1}}\circ \*V_{e_{m2}} \big)
\right] - \frac{\diag(\*\Lambda_k)}{\lambda}  &
\\ \nonumber
\frac{\partial Q}{\partial \*Y_k}  & = \sum^M_{m=1} p_{mk}  (\*U_{e_{m1}} + \*V_{e_{m2}} -  \frac{s_{uk}}{f_{uk}} -  \frac{s_{vk} -  e^{R_1{e_{m1}} + \*V_{e_{m1}}\*Y'_k} \*V_{e_{m1}}}{f_{vk} - e^{R_1{e_{m1}} + \*V_{e_{m1}}\*Y'_k}}) - \*Y_k &
\\ \nonumber
\frac{\partial Q}{\partial \beta_k}  & = \sum^M_{m=1} p_{mk} \left[ \frac{1}{a(\phi_k)}\left(w_m - \frac{\partial A(\eta)}{\partial \eta}\right) -
\frac{\frac{\partial}{\partial \eta_{e_{m1}e_{m2}k}} \Pr(w_m\neq 0|\eta_{e_{m1}e_{m2}k})}{\Pr(w_m\neq 0|\eta_{e_{m1}e_{m2}k})}
\right] &
\\ \nonumber
\frac{\partial Q}{\partial \phi_k} & = 
\sum_{m=1}^M p_{mk}\left[ \frac{\frac{\partial}{\partial \phi_k}h(w_m,\phi_k)}{h(w_m,\phi_k)} - \frac{\eta_{e_{m1}e_{m2}k}w_m - A(\eta)}{a^2(\phi_k)}\frac{d a(\phi_k)}{d\phi_k} 
 - \frac{\frac{\partial}{\partial \phi_k} \Pr(w_m\neq 0|\eta_{e_{m1}e_{m2}k})}{\Pr(w_m\neq 0|\eta_{e_{m1}e_{m2}k})} \right] - \frac{(\nu_0 + 1)\phi_k}{\nu_0\eta_0^2 + \phi_k^2} &
\end{align}

Where in the example of normal distribution, we have:
$$ \Pr(w \neq 0 | \eta_{ijk}) = 1, $$ 
$$ (\partial A(\eta_{ijk}))/(\partial \eta_{ijk}) = \eta_{ijk}, $$ 
$$ \frac{\frac{\partial}{\partial \eta_{ijk}} \Pr(w_m\neq 0|\eta_{ijk})}{\Pr(w_m\neq 0|\eta_{ijk})} = 0, $$ 
$$ \frac{\partial a(\phi_k)}{\partial \phi_k} = 2\phi_k, $$ 
$$  \frac{\partial log(h(w_m,\phi_k))}{\partial \phi_k} = \frac{w_m^2}{\phi_k^3} - \frac{1}{\phi_k} $$ 

Due to the potentially high computation cost, we introduce the following quantities to increase the estimation efficiency. It has been proven that, with these precomputed quantities, the time of computing these derivations can be reduced from $O_{(nM)}$ to $O_{(M+n)}$ \citep{sewell2021model}:

\begin{align*}
H_k &:= \sum_{m=1}^M\frac{p_{mk}}{f_{vk} - e^{R_{1e_{m1}} + \*V_{e_{m1}}\*Y_k'} }, & 
\\
s_{uk} & := \sum^n_{i=1}e^{S_{1i}+\*U_i\*Y'_k}\*U_i, &
\\
\mbox{and } s_{vk} &:= \sum^n_{i=1}e^{R_{1i}+\*V_i\*Y'_k}\*V_i. &
\end{align*}

\section{The analytical solutions}
\label{sec:appendixC}
The full conditionals of covariance-related variables $\{\widehat {\*\Sigma}_{SR1}, \widehat {\*\Sigma}_{SR2}, \widehat {\*\Sigma}_{UV}, \alpha_k, \lambda \}$ can be found as following:
\begin{align*}
    \*\Sigma_{SR1} | \*S_1, \*R_1 & \sim IW(\*\Psi_{nSR1}, \upsilon_{nSR1}), \\
    \*\Sigma_{SR_2} | \*S_2, \*R_2 & \sim IW(\*\Psi_{nSR2}, \upsilon_{nSR2}), \\
    \*\Sigma_{UV} | \*U, \*V & \sim IW(\*\Psi_{nUV}, \upsilon_{nUV}), \\
     \*\alpha | \cdot & \sim Dir(\alpha_0 + \sum_mp_{m1}, \alpha_0 + \sum_mp_{m2} , \cdots,  \alpha_0 + \sum_mp_{mk}), \\
     \lambda | \cdot & \sim \Gamma^{-1}\left(a_0 + \frac{Kp}{2}, b_0 + \frac{1}{2} \sum_k\|\diag(\Lambda_k)\|^2\right).
\end{align*}

with current estimations, $\{\*\Sigma_{SR_1}, \*\Sigma_{SR_2}, \*\Sigma_{UV}, \lambda, \*\alpha \} $ are from analytical solutions:
 $$\hat{\alpha_k} = \frac{\alpha_0 + \sum_mp_{mk} - 1}{K \alpha_0 + M - K},
\ \hat{\lambda} = \frac{b_0 + \frac{1}{2} \sum_k\|\diag(\Lambda_k)\|^2}{a_0 + \frac{Kp}{2} + 1}$$.
$$\widehat {\*\Sigma}_{SR1} = \frac{\*\Psi_{nSR1}}{\upsilon_{nSR1} + 3}, \ \widehat {\*\Sigma}_{SR2} = \frac{\*\Psi_{nSR2}}{\upsilon_{nSR2} + 3}, \  \widehat {\*\Sigma}_{UV} = \frac{\Psi_{nUV}}{\upsilon_{nUV} + 3}. $$
where: $\*\Psi_{nSR1} = \*\Psi_{0SR1} + \sum_{i = 1}^n \begin{pmatrix} S_{1i}^2 & S_{1i} R_{1i}\\ S_{1i} R_{1i}&R_{1i}^2 \end{pmatrix} $, and $\upsilon_{nSR1} = \upsilon_{0SR1} + n$; \\

$\*\Psi_{nSR2} = \*\Psi_{0SR2} + \sum_{i = 1}^n \begin{pmatrix} S_{2i}^2 & S_{2i} R_{2i}\\ S_{2i} R_{2i} & R_{2i}^2 \end{pmatrix} $, and $\upsilon_{SR2} = \upsilon_{0SR2} + n$; \\

$\*\Psi_{nUV} = \*\Psi_{0UV} + \sum_{i = 1}^n (\*U_i' \quad \*V_i')' (\*U_i \quad \*V_i) $, $\upsilon_{nUV} = \upsilon_{0UV} + np$.

\bibliographystyle{elsarticle-harv} 
\bibliography{cas-refs}

\begin{thebibliography}{30}
\expandafter\ifx\csname natexlab\endcsname\relax\def\natexlab#1{#1}\fi
\providecommand{\url}[1]{\texttt{#1}}
\providecommand{\href}[2]{#2}
\providecommand{\path}[1]{#1}
\providecommand{\DOIprefix}{doi:}
\providecommand{\ArXivprefix}{arXiv:}
\providecommand{\URLprefix}{URL: }
\providecommand{\Pubmedprefix}{pmid:}
\providecommand{\doi}[1]{\href{http://dx.doi.org/#1}{\path{#1}}}
\providecommand{\Pubmed}[1]{\href{pmid:#1}{\path{#1}}}
\providecommand{\bibinfo}[2]{#2}
\ifx\xfnm\relax \def\xfnm[#1]{\unskip,\space#1}\fi
\bibitem[{Aicher et~al.(2015)Aicher, Jacobs and Clauset}]{aicher2015learning}
\bibinfo{author}{Aicher, C.}, \bibinfo{author}{Jacobs, A.Z.}, \bibinfo{author}{Clauset, A.}, \bibinfo{year}{2015}.
\newblock \bibinfo{title}{Learning latent block structure in weighted networks}.
\newblock \bibinfo{journal}{Journal of Complex Networks} \bibinfo{volume}{3}, \bibinfo{pages}{221--248}.
\bibitem[{Amelio and Pizzuti(2014)}]{amelio2014overlapping}
\bibinfo{author}{Amelio, A.}, \bibinfo{author}{Pizzuti, C.}, \bibinfo{year}{2014}.
\newblock \bibinfo{title}{Overlapping community discovery methods: a survey}.
\newblock \bibinfo{journal}{Social networks: Analysis and case studies} , \bibinfo{pages}{105--125}.
\bibitem[{Banfield and Raftery(1993)}]{banfield1993model}
\bibinfo{author}{Banfield, J.D.}, \bibinfo{author}{Raftery, A.E.}, \bibinfo{year}{1993}.
\newblock \bibinfo{title}{Model-based gaussian and non-gaussian clustering}.
\newblock \bibinfo{journal}{Biometrics} , \bibinfo{pages}{803--821}.
\bibitem[{Ben-David and Haghtalab(2014)}]{ben2014clustering}
\bibinfo{author}{Ben-David, S.}, \bibinfo{author}{Haghtalab, N.}, \bibinfo{year}{2014}.
\newblock \bibinfo{title}{Clustering in the presence of background noise}, in: \bibinfo{booktitle}{International Conference on Machine Learning}, \bibinfo{organization}{PMLR}. pp. \bibinfo{pages}{280--288}.
\bibitem[{Berardo et~al.(2020)Berardo, Fischer and Hamilton}]{berardo2020collaborative}
\bibinfo{author}{Berardo, R.}, \bibinfo{author}{Fischer, M.}, \bibinfo{author}{Hamilton, M.}, \bibinfo{year}{2020}.
\newblock \bibinfo{title}{Collaborative governance and the challenges of network-based research}.
\newblock \bibinfo{journal}{The American Review of Public Administration} \bibinfo{volume}{50}, \bibinfo{pages}{898--913}.
\bibitem[{Bernardo et~al.(2003)Bernardo, Bayarri, Berger, Dawid, Heckerman, Smith, West et~al.}]{bernardo2003variational}
\bibinfo{author}{Bernardo, J.}, \bibinfo{author}{Bayarri, M.}, \bibinfo{author}{Berger, J.}, \bibinfo{author}{Dawid, A.}, \bibinfo{author}{Heckerman, D.}, \bibinfo{author}{Smith, A.}, \bibinfo{author}{West, M.}, et~al., \bibinfo{year}{2003}.
\newblock \bibinfo{title}{The variational bayesian em algorithm for incomplete data: with application to scoring graphical model structures}.
\newblock \bibinfo{journal}{Bayesian statistics} \bibinfo{volume}{7}, \bibinfo{pages}{210}.
\bibitem[{Biernacki et~al.(2000)Biernacki, Celeux and Govaert}]{biernacki2000assessing}
\bibinfo{author}{Biernacki, C.}, \bibinfo{author}{Celeux, G.}, \bibinfo{author}{Govaert, G.}, \bibinfo{year}{2000}.
\newblock \bibinfo{title}{Assessing a mixture model for clustering with the integrated completed likelihood}.
\newblock \bibinfo{journal}{IEEE transactions on pattern analysis and machine intelligence} \bibinfo{volume}{22}, \bibinfo{pages}{719--725}.
\bibitem[{Butts(2009)}]{butts2009revisiting}
\bibinfo{author}{Butts, C.T.}, \bibinfo{year}{2009}.
\newblock \bibinfo{title}{Revisiting the foundations of network analysis}.
\newblock \bibinfo{journal}{science} \bibinfo{volume}{325}, \bibinfo{pages}{414--416}.
\bibitem[{Danon et~al.(2005)Danon, Diaz-Guilera, Duch and Arenas}]{danon2005comparing}
\bibinfo{author}{Danon, L.}, \bibinfo{author}{Diaz-Guilera, A.}, \bibinfo{author}{Duch, J.}, \bibinfo{author}{Arenas, A.}, \bibinfo{year}{2005}.
\newblock \bibinfo{title}{Comparing community structure identification}.
\newblock \bibinfo{journal}{Journal of statistical mechanics: Theory and experiment} \bibinfo{volume}{2005}, \bibinfo{pages}{P09008}.
\bibitem[{Dempster et~al.(1977)Dempster, Laird and Rubin}]{dempster1977maximum}
\bibinfo{author}{Dempster, A.P.}, \bibinfo{author}{Laird, N.M.}, \bibinfo{author}{Rubin, D.B.}, \bibinfo{year}{1977}.
\newblock \bibinfo{title}{Maximum likelihood from incomplete data via the em algorithm}.
\newblock \bibinfo{journal}{Journal of the royal statistical society: series B (methodological)} \bibinfo{volume}{39}, \bibinfo{pages}{1--22}.
\bibitem[{Donker et~al.(2010)Donker, Wallinga and Grundmann}]{donker2010patient}
\bibinfo{author}{Donker, T.}, \bibinfo{author}{Wallinga, J.}, \bibinfo{author}{Grundmann, H.}, \bibinfo{year}{2010}.
\newblock \bibinfo{title}{Patient referral patterns and the spread of hospital-acquired infections through national health care networks}.
\newblock \bibinfo{journal}{PLoS computational biology} \bibinfo{volume}{6}, \bibinfo{pages}{e1000715}.
\bibitem[{Eddelbuettel and Sanderson(2014)}]{RcppArmadillo2014}
\bibinfo{author}{Eddelbuettel, D.}, \bibinfo{author}{Sanderson, C.}, \bibinfo{year}{2014}.
\newblock \bibinfo{title}{Rcpparmadillo: Accelerating r with high-performance c++ linear algebra}.
\newblock \bibinfo{journal}{Computational Statistics and Data Analysis} \bibinfo{volume}{71}, \bibinfo{pages}{1054--1063}.
\newblock \DOIprefix\doi{10.1016/j.csda.2013.02.005}.
\bibitem[{Evans and Lambiotte(2009)}]{evans2009line}
\bibinfo{author}{Evans, T.S.}, \bibinfo{author}{Lambiotte, R.}, \bibinfo{year}{2009}.
\newblock \bibinfo{title}{Line graphs, link partitions, and overlapping communities}.
\newblock \bibinfo{journal}{Physical review E} \bibinfo{volume}{80}, \bibinfo{pages}{016105}.
\bibitem[{Evans and Lambiotte(2010)}]{evans2010line}
\bibinfo{author}{Evans, T.S.}, \bibinfo{author}{Lambiotte, R.}, \bibinfo{year}{2010}.
\newblock \bibinfo{title}{Line graphs of weighted networks for overlapping communities}.
\newblock \bibinfo{journal}{The European Physical Journal B} \bibinfo{volume}{77}, \bibinfo{pages}{265--272}.
\bibitem[{Handcock et~al.(2007)Handcock, Raftery and Tantrum}]{handcock2007model}
\bibinfo{author}{Handcock, M.S.}, \bibinfo{author}{Raftery, A.E.}, \bibinfo{author}{Tantrum, J.M.}, \bibinfo{year}{2007}.
\newblock \bibinfo{title}{Model-based clustering for social networks}.
\newblock \bibinfo{journal}{Journal of the Royal Statistical Society: Series A (Statistics in Society)} \bibinfo{volume}{170}, \bibinfo{pages}{301--354}.
\bibitem[{Hoff et~al.(2002)Hoff, Raftery and Handcock}]{hoff2002latent}
\bibinfo{author}{Hoff, P.D.}, \bibinfo{author}{Raftery, A.E.}, \bibinfo{author}{Handcock, M.S.}, \bibinfo{year}{2002}.
\newblock \bibinfo{title}{Latent space approaches to social network analysis}.
\newblock \bibinfo{journal}{Journal of the american Statistical association} \bibinfo{volume}{97}, \bibinfo{pages}{1090--1098}.
\bibitem[{Kalinka and Tomancak(2011)}]{linkcomm2011}
\bibinfo{author}{Kalinka, A.T.}, \bibinfo{author}{Tomancak, P.}, \bibinfo{year}{2011}.
\newblock \bibinfo{title}{linkcomm: an r package for the generation, visualization, and analysis of link communities in networks of arbitrary size and type}.
\newblock \bibinfo{journal}{Bioinformatics} \bibinfo{volume}{27}, \bibinfo{pages}{2011--2012}.
\newblock \DOIprefix\doi{10.1093/bioinformatics/btr311}.
\bibitem[{Karrer and Newman(2011)}]{PhysRevE.83.016107}
\bibinfo{author}{Karrer, B.}, \bibinfo{author}{Newman, M.E.J.}, \bibinfo{year}{2011}.
\newblock \bibinfo{title}{Stochastic blockmodels and community structure in networks}.
\newblock \bibinfo{journal}{Phys. Rev. E} \bibinfo{volume}{83}, \bibinfo{pages}{016107}.
\newblock \URLprefix \url{https://link.aps.org/doi/10.1103/PhysRevE.83.016107}, \DOIprefix\doi{10.1103/PhysRevE.83.016107}.
\bibitem[{Kernighan and Lin(1970)}]{kernighan1970efficient}
\bibinfo{author}{Kernighan, B.W.}, \bibinfo{author}{Lin, S.}, \bibinfo{year}{1970}.
\newblock \bibinfo{title}{An efficient heuristic procedure for partitioning graphs}.
\newblock \bibinfo{journal}{The Bell system technical journal} \bibinfo{volume}{49}, \bibinfo{pages}{291--307}.
\bibitem[{Khan and Niazi(2017)}]{khan2017network}
\bibinfo{author}{Khan, B.S.}, \bibinfo{author}{Niazi, M.A.}, \bibinfo{year}{2017}.
\newblock \bibinfo{title}{Network community detection: A review and visual survey}.
\newblock \bibinfo{journal}{arXiv preprint arXiv:1708.00977} .
\bibitem[{Newman and Girvan(2004)}]{newman2004finding}
\bibinfo{author}{Newman, M.E.}, \bibinfo{author}{Girvan, M.}, \bibinfo{year}{2004}.
\newblock \bibinfo{title}{Finding and evaluating community structure in networks}.
\newblock \bibinfo{journal}{Physical review E} \bibinfo{volume}{69}, \bibinfo{pages}{026113}.
\bibitem[{Ng et~al.(2001)Ng, Jordan and Weiss}]{ng2001spectral}
\bibinfo{author}{Ng, A.}, \bibinfo{author}{Jordan, M.}, \bibinfo{author}{Weiss, Y.}, \bibinfo{year}{2001}.
\newblock \bibinfo{title}{On spectral clustering: Analysis and an algorithm}.
\newblock \bibinfo{journal}{Advances in neural information processing systems} \bibinfo{volume}{14}.
\bibitem[{Nowicki and Snijders(2001)}]{nowicki2001estimation}
\bibinfo{author}{Nowicki, K.}, \bibinfo{author}{Snijders, T.A.B.}, \bibinfo{year}{2001}.
\newblock \bibinfo{title}{Estimation and prediction for stochastic blockstructures}.
\newblock \bibinfo{journal}{Journal of the American statistical association} \bibinfo{volume}{96}, \bibinfo{pages}{1077--1087}.
\bibitem[{Opsahl and Panzarasa(2009)}]{opsahl2009clustering}
\bibinfo{author}{Opsahl, T.}, \bibinfo{author}{Panzarasa, P.}, \bibinfo{year}{2009}.
\newblock \bibinfo{title}{Clustering in weighted networks}.
\newblock \bibinfo{journal}{Social networks} \bibinfo{volume}{31}, \bibinfo{pages}{155--163}.
\bibitem[{Pham and Sewell(2024)}]{pham2024automated}
\bibinfo{author}{Pham, H.T.}, \bibinfo{author}{Sewell, D.K.}, \bibinfo{year}{2024}.
\newblock \bibinfo{title}{Automated detection of edge clusters via an overfitted mixture prior}.
\newblock \bibinfo{journal}{Network Science} , \bibinfo{pages}{1--19}.
\bibitem[{{R Core Team}(2023)}]{Ritself2023}
\bibinfo{author}{{R Core Team}}, \bibinfo{year}{2023}.
\newblock \bibinfo{title}{R: A Language and Environment for Statistical Computing}.
\newblock \bibinfo{organization}{R Foundation for Statistical Computing}. \bibinfo{address}{Vienna, Austria}.
\newblock \URLprefix \url{https://www.R-project.org/}.
\bibitem[{Sewell(2021)}]{sewell2021model}
\bibinfo{author}{Sewell, D.K.}, \bibinfo{year}{2021}.
\newblock \bibinfo{title}{Model-based edge clustering}.
\newblock \bibinfo{journal}{Journal of Computational and Graphical Statistics} \bibinfo{volume}{30}, \bibinfo{pages}{390--405}.
\bibitem[{Tian et~al.(2023)Tian, Lubberts and Weber}]{tian2023mixed}
\bibinfo{author}{Tian, Y.}, \bibinfo{author}{Lubberts, Z.}, \bibinfo{author}{Weber, M.}, \bibinfo{year}{2023}.
\newblock \bibinfo{title}{Mixed-membership community detection via line graph curvature}, in: \bibinfo{booktitle}{NeurIPS Workshop on Symmetry and Geometry in Neural Representations}, \bibinfo{organization}{PMLR}. pp. \bibinfo{pages}{219--233}.
\bibitem[{Von~Luxburg(2007)}]{von2007tutorial}
\bibinfo{author}{Von~Luxburg, U.}, \bibinfo{year}{2007}.
\newblock \bibinfo{title}{A tutorial on spectral clustering}.
\newblock \bibinfo{journal}{Statistics and computing} \bibinfo{volume}{17}, \bibinfo{pages}{395--416}.
\bibitem[{Yoshida(2013)}]{yoshida2013weighted}
\bibinfo{author}{Yoshida, T.}, \bibinfo{year}{2013}.
\newblock \bibinfo{title}{Weighted line graphs for overlapping community discovery}.
\newblock \bibinfo{journal}{Social Network Analysis and Mining} \bibinfo{volume}{3}, \bibinfo{pages}{1001--1013}.

\end{thebibliography}

\end{document}